\documentclass[aps,prd,onecolumn,groupedaddress,showpacs,nofootinbib,amssymb]{revtex4}
\usepackage[dvips]{graphicx}
\usepackage{amssymb}
\usepackage{amsmath}
\usepackage{graphicx,,color}
\usepackage{amsfonts}
\usepackage{bm}
\usepackage{cancel}
\usepackage{comment}

\newcommand\be{\begin{equation}}
\newcommand\ee{\end{equation}}

\allowdisplaybreaks[4]

\begin{document}

\title{Towards Model Agnostic $F(R)$ Gravity Inflation}
\author{V.K. Oikonomou,$^{1,2}$}\email{voikonomou@gapps.auth.gr;v.k.oikonomou1979@gmail.com}
\affiliation{$^{1)}$Department of Physics, Aristotle University of
Thessaloniki, Thessaloniki 54124, Greece \\ $^{2)}$L.N. Gumilyov
Eurasian National University - Astana, 010008, Kazakhstan}

\tolerance=5000

\begin{abstract}
In this work we construct a formalism that can reveal the general
characteristics of classes of viable $F(R)$ inflationary theories.
The assumptions we make is that the slow-roll era occurs, and that
the de Sitter scalaron mass $m^2(R)$ of the $F(R)$ gravity is
positive or zero, for both the inflationary and late-time quasi de
Sitter eras, a necessary condition for the stability of the de
Sitter spacetime. In addition, we require that the de Sitter
scalaron mass is also a monotonically increasing function of the
Ricci scalar, or it has an extremum. Also the $F(R)$ gravity
function is required to depend on the two known fundamental scales
in cosmology, the cosmological constant $\Lambda$ and the mass
scale $m_s^2=\frac{\kappa^2 \rho_m^{(0)}}{3}$, with $\rho_m^{(0)}$
denoting the energy density of the cold dark matter at the present
epoch, that is $F(R)=F(R,\Lambda,m_s^2)$. Using these general
assumptions we provide the general features of viable classes of
$F(R)$ gravity inflationary theories which remarkably can also
simultaneously describe successfully the dark energy era. This
unique feature of a unified description of the dark energy and
inflationary eras stems from the requirement of the monotonicity
of the de Sitter scalaron mass $m^2(R)$. These viable classes are
either deformations of the $R^2$ model or $\alpha$-attractors type
theories. The analysis of the viability of a general $F(R)$
gravity inflationary theory is reduced in evaluating the parameter
$x=\frac{R F_{RRR}}{F_{RR}}$ and the first slow-roll index of the
theory, either numerically or approximately. We also disentangle
the power-law $F(R)$ gravities from power-law evolution and we
show that power-law $F(R)$ gravities can be viable theories of
inflation, for appropriate values of the power-law exponent.
Finally we highlight the phenomenological importance of
exponential deformations of the $R^2$ model of the form
$F(R)=R+\frac{R^2}{M^2}+\lambda
R\,e^{\epsilon\left(\frac{\Lambda}{R}\right)^{\beta}}+\lambda
\Lambda n \epsilon$, which emerge naturally as viable inflationary
models which also describe successfully the dark energy era.
\end{abstract}


\maketitle

\section{Introduction}

Undoubtedly the post-Planck era of our Universe is one of the most
mysterious cosmological eras that can be, hopefully,
observationally verified. The prominent candidate theory for the
description of the post-Planck four dimensional classical Universe
is inflation
\cite{inflation1,inflation2,inflation3,inflation4,inflation5}.
Inflation by itself as a theoretical construction of the human
mind is remarkable since it solves in an elegant way all the
problems of the classical Big Bang theory. Apart from the elegant
theoretical description that inflation offers to the post-quantum
era of our Universe, the future experiments in the night sky aim
to observationally verify this mysterious epoch of our Universe.
Indeed, the Simons observatory \cite{SimonsObservatory:2019qwx}
and the stage 4 Cosmic Microwave Background (CMB) experiments
\cite{CMB-S4:2016ple}, if hopefully these commence, aim to provide
a direct detection of the curl models of inflation, so-called
$B$-modes \cite{Kamionkowski:2015yta}. The detection of the
$B$-modes in the CMB, will verify directly the existence of tensor
perturbations in the CMB, a smoking gun for inflation. Now apart
from the near future CMB experiments, the future gravitational
wave experiments offer the fascinating possibility of detecting a
stochastic gravitational wave background that can be generated by
some inflationary theories
\cite{Hild:2010id,Baker:2019nia,Smith:2019wny,Crowder:2005nr,Smith:2016jqs,Seto:2001qf,Kawamura:2020pcg,Bull:2018lat,LISACosmologyWorkingGroup:2022jok}.
Even in 2023, the existence of a stochastic gravitational wave
background has been verified
\cite{nanograv,EPTA:2023fyk,Reardon:2023gzh,Xu:2023wog}, however
this tensor perturbation background is highly unlikely to be the
effect of an inflationary era by itself solely
\cite{sunnynew,Oikonomou:2023qfz,Vagnozzi:2020gtf}. Inflation can
be realized in a customary way in the context of general
relativity by using a single scalar field theory, but it can also
be realized in a geometric way, by modifying Einstein's
gravitational theory \cite{reviews1,reviews2,reviews3}. Both the
scalar field and modified gravity description have their own
inherent appeal, for different reasons. The scalar theory utilizes
a scalar field, the so-called inflaton, which is motivated by the
existence of the Higgs field and due to the fact that scalar
fields are remnants of the possible ultra-violet extension of the
Standard Model, namely remnants of string theory. On the other
hand, the downside of the single scalar field description of the
Universe is that the inflaton has to have too many couplings to
the Standard Model particles in order to reheat the Universe. Thus
unless the inflaton is the Higgs itself \cite{Bezrukov:2007ep},
the single scalar field description can be somewhat artificial.
The modified gravity approach on the other had utilizes a
geometric description for both the inflationary era and the
reheating, and is again motivated by string theory, since higher
curvature terms often occur in Einstein's gravity as remnants of
string theory. There are various modified gravity theories that
can realize an inflationary era in a successful way, but the most
prominent of these theories is $F(R)$ gravity. The reason why
$F(R)$ gravity enjoys an elevated role among other modified
gravities is mainly its simplicity, and also the fact that, from a
mathematical point of view, the curvature corrections are
naturally simplest based on a theory composed on the vector bundle
of general relativity with the local principal bundle having the
transformation group $GL(4,R)$. In terms of connections, thus in
terms of wedge products, $F(R)$ gravity emerges as the simplest
generalization of Einstein's gravity on a four dimensional
manifold. There are also other reasons for discussing $F(R)$
gravity corrections in Einstein-Hilbert gravity, with the most
prominent being the description of the dark energy era. With the
publication of the 2024 DESI observational data
\cite{DESI:2024mwx}, which point out that the dark energy is
dynamical, $F(R)$ gravity realizations are quite timely and
popular. This fact has significantly been amplified this year,
with the DESI 2025 data release \cite{DESI:2025zgx} which indicate
that dark energy is dynamical at very late times up to $4.2\sigma$
statistical confidence. Moreover the DESI 2025 data indicate a
transition of the dark energy equation of state (EoS) from a
phantom value $w<-1$ to a quintessential value $w>-1$ at very late
times. This clearly indicates that general relativity is
challenged at late times, because a phantom regime realization in
the context of general relativity would require tachyon fields and
also the transition itself could be difficult to achieve. In the
context of $F(R)$ gravity, such cosmological scenarios are easy to
realize without invoking exotic components. There exists a vast
literature on both inflationary and dark energy aspects of $F(R)$
gravity, and for a mainstream of articles on this timely topic see
Refs.
\cite{Nojiri:2003ft,Capozziello:2005ku,Capozziello:2004vh,Capozziello:2018ddp,Hwang:2001pu,Cognola:2005de,Nojiri:2006gh,Song:2006ej,Capozziello:2008qc,Bean:2006up,Capozziello:2012ie,Faulkner:2006ub,Olmo:2006eh,Sawicki:2007tf,Faraoni:2007yn,Carloni:2007yv,
Nojiri:2007as,Capozziello:2007ms,Deruelle:2007pt,Appleby:2008tv,Dunsby:2010wg,Odintsov:2020nwm,Odintsov:2019mlf,Odintsov:2019evb,Oikonomou:2020oex,Oikonomou:2020qah,Huang:2013hsb,Capozziello:2012ie,Berry:2011pb,Bonanno:2010bt,Gannouji:2008wt,Oyaizu:2008sr,Oyaizu:2008tb,Brax:2008hh,Cognola:2007zu,Boehmer:2007glt,Boehmer:2007kx,
deSouza:2007zpn,Song:2007da,Brookfield:2006mq,delaCruz-Dombriz:2006kob,Myrzakulov:2015qaa,Achitouv:2015yha,Kopp:2013lea,Paliathanasis:2017apr,Leon:2022dwd}
and references therein. One appealing perspective in the context
of $F(R)$ gravity is to describe in a unified way inflation and
the dark energy era. This line of research was firstly realized in
the pioneer work \cite{Nojiri:2003ft} and later developments were
given in Refs.
\cite{Nojiri:2006gh,Nojiri:2007as,Appleby:2008tv,Odintsov:2019evb,Oikonomou:2020oex,Oikonomou:2020qah}.
Most of the known unified descriptions of $F(R)$ inflation and
dark energy, mimic the $\Lambda$-Cold-Dark-Matter model
($\Lambda$CDM) at late times.

However, only a handful of $F(R)$ gravity models can be solved
analytically. The inflationary era can be realized by a quasi-de
Sitter evolution, and the only model that can yield analytically a
quasi-de Sitter evolution is the $R^2$ model
\cite{Starobinsky:1980te}. Apart from that it is quite hard to
solve and study distinct $F(R)$ gravity models. In this article we
aim to provide a general and model agnostic method in order to
decide whether a given $F(R)$ gravity can produce a viable
inflationary era. Our approach is simple, and we assume that the
$F(R)$ gravity function depends on the Ricci scalar, and the only
two known fundamental scales in cosmology, the cosmological
constant and the mass scale $m_s^2$, where $m_s^2=\frac{\kappa^2
\rho_m^{(0)}}{3}$, with $\rho_m^{(0)}$ denoting the energy density
of the cold dark matter at present time. Remarkably, with this
assumption, we managed to find several classes of viable
inflationary models, that can also produce a viable dark energy
era. Thus, by trying to find viable inflationary models, we
provide a self-efficient technique to also find viable dark energy
models, not based on phenomenology, by adding by hand terms, but
via a formal procedure aimed for inflationary dynamics. This is
the first time that such a development has appeared in the
literature. Regarding the viable inflationary era, by using only
the assumption of a slow-roll era $\dot{H}\ll H^2$, and also that
the first slow-roll index $\epsilon_1$ is non-constant, that is
$\dot{\epsilon}_1\neq 0$, we produce a formalism for studying in a
compact way $F(R)$ gravity inflationary dynamics. The first steps
of this part of the analysis was also developed in Ref.
\cite{Odintsov:2020thl}. As we demonstrate, the scalar spectral
index in the large curvature slow-roll regime takes the form,
$$
n_s-1=-4\epsilon_1+x\epsilon_1\, ,
$$
and the tensor-to-scalar ratio is,
$$
r\simeq \frac{48 (1-n_s)^2}{(4-x)^2}\, .
$$
The parameter $x$ defined as,
$$
x=\frac{4 F_{RRR}\,R}{F_{RR}}\, ,
$$
will prove to play a fundamental role in our analysis. Now one
major assumption in this work, which is theoretically strongly
motivated in the line of research of a unified inflation and dark
energy description, is that we will assume that the de Sitter
scalaron mass of $F(R)$ gravity is a monotonically increasing
function of the Ricci scalar or has an extremum, in the large
curvature slow-roll regime. The de Sitter scalaron mass is defined
as,
$$
m^2(R)=\frac{1}{3}\left(-R+\frac{F_R}{F_{RR}} \right)\, ,
$$
or in terms of the variable $y$,
$$
m^2=\frac{R}{3}\left(-1+\frac{1}{y} \right)\, .
$$
with $y$,
$$
y=\frac{R\,F_{RR}}{F_R}\, .
$$
Thus the main assumption is that the function,
$$
m^2(R)=\frac{1}{3}\left(-1+\frac{F_R }{F_{RR}R} \right)
$$
is a monotonically increasing function of $R$, or has an extremum.
This has dramatic consequences for the allowed $F(R)$ gravities.
Remarkably, as it proves in the end, for the $F(R)$ gravities we
found, the scalaron mass is small at small curvatures and large at
large curvatures. This is theoretically motivated by the late-time
behavior of the scalaron mass, and we shall further elaborate on
this at a later point. So we require,
$$
\frac{\partial m^2}{\partial R}\geq 0\, ,
$$
or in terms of the function $F(R)$,
$$
\frac{\partial m^2}{\partial
R}=-\frac{1}{12}\frac{F_R}{R\,F_{RR}}\,\frac{4\,R\,F_{RRR}}{F_{RR}}\geq
0 \, ,
$$
or equivalently,
$$
\frac{\partial m^2}{\partial R}=-\frac{1}{12}\frac{x}{y}\geq 0 \,
.
$$
Also the scalaron mass is demanded to be positive or zero, for
both the inflationary and late-time evolution eras, in order to
ensure stability of de Sitter spacetime, thus,
$$
0< y \leq 1\, ,
$$
so the two requirements can be met only when,
$$
x\leq 0,\,\,\,0< y\leq 1
$$
Thus viable inflationary theories, which can also be consistent at
late times, must yield $x\leq 0$ and $0<y\leq1$ in the large
curvature slow-roll regime and also the first slow-roll index must
be appropriately small at first horizon crossing. From the form of
the spectral index in terms of $x$ and the first slow-roll index,
it proves that most viable inflation scenarios are found for
$-1\leq x\leq 0$, if one assumes that the first slow-roll index
does not take extremely small values. These are the features of
all the viable $F(R)$ gravities which can also be theoretically
consistent at late times. We examine several classes of viable
inflationary theories and provide the general features of these
viable classes of models. As it occurs, the viable models are
classified in two main classes, either $R^2$ deformations, or
theories that lead to $\alpha$-attractor-like
\cite{alpha1,alpha2,alpha3,alpha4,alpha5,alpha6,alpha7,alpha8,alpha9,alpha10,alpha10a,alpha11,alpha12,vernov}
behavior during inflation. We also study several cases for which
non-viable inflationary theories are obtained. Another important
task which we perform in this work is the study of power-law type
$F(R)$ gravities. These theories result to a constant $x$
parameter, and in the literature these theories are linked to
power-law evolution. As we show, this is not true, and we
disentangle the power-law evolution from power-law $F(R)$
gravities. As we show, power-law gravities can be viable theories,
and we also provide an estimate of the first slow-roll index for
$F(R)$ gravities, which can serve as an estimate for the order of
magnitude of $\epsilon_1$, namely, the formula,
$$
\epsilon_1\sim \frac{2 F-F_R R}{2F_{RR}R^2}\, .
$$
Thus our method makes the study of inflationary $F(R)$ gravity
theories quite easy, since we only need to find the parameter $x$
and the first slow-roll index for the analysis. We also provide a
method for analyzing inflationary $F(R)$ gravity theories and we
stress the need for numerical analysis for the first slow-roll
index solely, in the case that accuracy is needed for a
potentially viable model. Finally, we demonstrate that the viable
$F(R)$ gravity models which are primordially  exponential $R^2$
model deformations, also lead to a viable dark energy era. This is
a remarkable result, since our analysis focused on the
inflationary era, but the resulting $F(R)$ gravities are also
excellent models for the dark energy era. The reason behind this
unified description of inflation and the dark energy era is the
fact that we demanded the scalaron mass to be a monotonically
increasing function of the curvature, and also due to the fact
that we demanded the $F(R)$ function to depend on the only known
mass scales in cosmology, the cosmological constant and the scale
$m_s^2=\frac{\kappa^2 \rho_m^{(0)}}{3}$. Finally, we highlight a
successful class of viable $F(R)$ gravity models which are able to
unify inflation with dark energy and naturally emerge from our
formalism. These are exponential deformations of the $R^2$ model,
of the form,
$$
F(R)=R+\frac{R^2}{M^2}+\lambda
R\,e^{\epsilon\left(\frac{\Lambda}{R}\right)^{\beta}}+\lambda
\Lambda n \epsilon
$$
with $\epsilon$, $\lambda$, $\beta$ and $n$ are dimensionless
parameters. These models yield an $R^2$ inflationary phenomenology
and at late times they produce a viable dark energy era and all
these models stem naturally by the formalism developed in this
paper.

Before starting our analysis, let us fix the background metric
which shall be used in this paper, and we assume that it is a flat
Friedmann-Robertson-Walker (FRW) metric with line element,
\begin{equation}
\label{JGRG14} ds^2 = - dt^2 + a(t)^2 \sum_{i=1,2,3}
\left(dx^i\right)^2\, ,
\end{equation}
with $a(t)$ being the scale factor and the Hubble rate is
$H=\frac{\dot{a}}{a}$.

\section{$F(R)$ Gravity Inflation and its Model Agnostic Formulation}

In this section we shall introduce the basic formalism for
studying $F(R)$ gravity inflation in a model agnostic approach. We
shall explain the most fundamental features of $F(R)$ gravity
inflation in the slow-roll regime and discuss the basic features
of a viable $F(R)$ gravity theory. This formalism will be used in
the next sections to reveal the features of $F(R)$ gravity
inflation without using a specific model.

\subsection{General Consideration for the $F(R)$ Gravity Action: Relevant Scales from Fundamental Physics and Viable $F(R)$ Gravity Constraints}

Let us start with the $F(R)$ gravity action, by considering the
general features that the $F(R)$ gravity function will posses. The
$F(R)$ gravity action in the absence of any matter fluids, will
have the general form,
\begin{equation}\label{FRgravitygeneralaction}
\mathcal{S}=\frac{1}{2\kappa^2}\int \mathrm{d}^4x\sqrt{-g}F(R)\, ,
\end{equation}
where $\kappa^2=8\pi G=\frac{1}{M_p^2}$, with $M_p$ being the
reduced Planck mass, $G$ is Newton's constant. Note that the
difference between the Planck mass $M_P$ and the reduced Planck
mass $M_p$ is a factor $4\pi$. Specifically, the Planck mass is
defined as,
\[
M_{P} = \sqrt{\frac{\hbar c}{G}} \approx 1.22 \times
10^{19}~\text{GeV}
\]
This is the mass scale where gravitational interactions become
strong and quantum gravitational effects become important. Also
the reduced Planck mass is,
\[
M_{p} = \frac{M_P}{\sqrt{8\pi}} = \sqrt{\frac{\hbar c}{8\pi G}}
\approx 2.43 \times 10^{18}~\text{GeV}
\]
This reduced Planck mass is used more often in modern field theory
and cosmology, since it simplifies the pre-factor of the
Einstein-Hilbert action, making many equations elegant. Thus one
must determine the functional form of $F(R)$ gravity in order to
perform the calculations for inflation or dark energy. But let us
start from the fundamental features that the $F(R)$ gravity
function will have. Basically, an $F(R)$ gravity is a
generalization of the Einstein-Hilbert gravity, and it is a higher
derivative theory. This theory must be somehow a remnant of the
quantum epoch of the Universe, which remained active after our
Universe left the quantum epoch and entered its classical four
dimensional epoch. It is natural to think that if this quantum
originating $F(R)$ gravity indeed exists, then it must somehow be
active during the whole evolution of our Universe, and not for
only one epoch, for instance the inflationary era or the dark
energy era. Thus the $F(R)$ gravity function should describe the
whole Universe evolution in a unified way. To date we have some
standard $F(R)$ gravity descriptions for inflation, like the $R^2$
model, or the dark energy epoch, see for example the models
developed in \cite{Oikonomou:2022wuk}. But a formally developed
unique and unified description of both inflation and dark energy
does not exist to date, although phenomenologically engineered
models exist in the literature. Our aim in this paper is to find a
formal way to connect the inflationary epoch with the dark energy
epoch, within the same $F(R)$ framework. Let us start with the
function $F(R)$ and think what constants and fundamental scales
will it contain. From cosmology, there are two mass scales that
must be somehow contained in the $F(R)$ gravity action. These are,
the cosmological constant $\Lambda$ and also the mass scale
$m_s^2=\frac{\kappa^2 \rho_m^{(0)}}{3}=H_0^2 \Omega_m=1.37 \times
10^{-67} eV^2$, where $\rho_m^{(0)}$ denotes the energy density of
the cold dark matter at the present epoch, with
$m_s^2=\frac{\kappa^2 \rho_m^{(0)}}{3}=H_0^2 \Omega_m=1.37 \times
10^{-67} eV^2$, and $H_0$ is the Hubble rate of the Universe at
present time. Thus one naturally expects that the $F(R)$ gravity
function will be of the general form,
\begin{equation}\label{FRgravitygeneralactionmainequationwithconstants}
\mathcal{S}=\frac{1}{2\kappa^2}\int
\mathrm{d}^4x\sqrt{-g}F(R,\Lambda,m_s^2)\, .
\end{equation}
In some way, the mass scales $\Lambda$, $m_s^2$ must be present in
function $F(R)$, if it genuinely describes nature from the
inflationary epoch to the dark energy epoch. Now there exist
several viability criteria for the functional form of $F(R)$
gravity, having to do with local solar system tests and also for
theoretical reasons \cite{reviews1,reviews2,reviews3}. Let us
quote the viability criteria here, for details see
\cite{reviews1,reviews2,reviews3}. The viability criteria are:
\begin{equation}\label{criterion1}
F_R>0
\end{equation}
where $F_R=\frac{\partial F}{\partial R}$, in order to avoid
anti-gravity, also,
\begin{equation}\label{criterion2}
F_{RR}>0
\end{equation}
where $F_{RR}=\frac{\partial^2 F}{\partial R^2}$, which is
required for the compatibility of the $F(R)$ gravity with local
solar system tests, and also for the occurrence of a successful
matter domination epoch and finally for the stability of the
cosmological perturbations. Finally, in order for a stable de
Sitter point exists as a solution, for both the inflationary
regime and the late-time regime, one must always have,
\begin{equation}\label{criterion3}
0< y \leq 1\, ,
\end{equation}
where $y$ is defined to be,
\begin{equation}\label{yparameterdefinition}
y=\frac{R\,F_{RR}}{F_R}\, .
\end{equation}
The de Sitter existence criterion is easily derived by perturbing
the field equations for a FRW spacetime, and specifically, if
$R=R_0+G(R)$, where $R_0$ is the scalar curvature of the de Sitter
point, the scalaron field in the Einstein frame obeys the
equation,
\begin{equation}\label{scalaronequation}
\square G+m^2 G=0\, ,
\end{equation}
with the scalaron mass being \cite{Muller:1987hp},
\begin{equation}\label{scalaronmassinitial}
m^2=\frac{1}{3}\left(-R+\frac{F_R}{F_{RR}} \right)\, ,
\end{equation}
or in terms of the variable $y$, the scalaron mass is written as
follows,
\begin{equation}\label{scalaronmassfinal}
m^2=\frac{R}{3}\left(-1+\frac{1}{y} \right)\, .
\end{equation}
Thus the scalaron mass is always positive or zero when the
condition (\ref{criterion3}) holds true. This requirement, also
constrains the first derivative of the scalaron mass with respect
to the Ricci scalar, since the scalaron mass must always be
positive or zero, and if the derivative of $m^2(R)$ is positive or
zero, the scalaron mass decreases as the curvature decreases and
the conversely, the scalaron mass should increase as the curvature
increases, or the scalaron mass has an extremum in the case the
derivative is zero. This will prove to be very valuable, as we
show later on in this section.

Having discussed the important features of the $F(R)$ gravity
function, let us proceed in formalizing the $F(R)$ gravity
inflation, without determining the $F(R)$ gravity function.

\subsection{Model Independent $F(R)$ Gravity Inflation}

Let us now review the formalism for the model agnostic $F(R)$
gravity inflation. A brief introduction to this approach was given
in Ref. \cite{Odintsov:2020thl} but this approach was an
introduction to the more focused and motivated approach of the
present article. Consider $F(R)$ gravity in vacuum, and thus the
action is given by Eq. (\ref{FRgravitygeneralaction}). We can
obtain the field equations in the metric formalism by varying the
gravitational action (\ref{FRgravitygeneralaction}) with respect
to the metric, thus the field equations read,
\begin{equation}\label{eqnmotion}
F_R(R)R_{\mu \nu}(g)-\frac{1}{2}F(R)g_{\mu
\nu}-\nabla_{\mu}\nabla_{\nu}F_R(R)+g_{\mu \nu}\square F_R(R)=0\,
,
\end{equation}
where recall that $F_R=\frac{\mathrm{d}F}{\mathrm{d}R}$. Eq.
(\ref{eqnmotion}) can be rewritten as follows,
\begin{align}\label{modifiedeinsteineqns}
R_{\mu \nu}-\frac{1}{2}Rg_{\mu
\nu}=\frac{\kappa^2}{F_R(R)}\Big{(}T_{\mu
\nu}+\frac{1}{\kappa^2}\Big{(}\frac{F(R)-RF_R(R)}{2}g_{\mu
\nu}+\nabla_{\mu}\nabla_{\nu}F_R(R)-g_{\mu \nu}\square
F_R(R)\Big{)}\Big{)}\, .
\end{align}
For the FRW metric of Eq. (\ref{JGRG14}), the field equations
acquire the following form,
\begin{align}
\label{JGRG15} 0 =& -\frac{F(R)}{2} + 3\left(H^2 + \dot H\right)
F_R(R) - 18 \left( 4H^2 \dot H + H \ddot H\right) F_{RR}(R)\, ,\\
\label{Cr4b} 0 =& \frac{F(R)}{2} - \left(\dot H +
3H^2\right)F_R(R) + 6 \left( 8H^2 \dot H + 4 {\dot H}^2 + 6 H
\ddot H + \dddot H\right) F_{RR}(R) + 36\left( 4H\dot H + \ddot
H\right)^2 F_{RRR} \, ,
\end{align}
where $F_{RR}=\frac{\mathrm{d}^2F}{\mathrm{d}R^2}$, and also
$F_{RRR}=\frac{\mathrm{d}^3F}{\mathrm{d}R^3}$. Furthermore $H$
denotes the Hubble rate and also $R$ denotes the Ricci scalar,
which for the FRW metric takes the form,
\begin{equation}\label{ricciscalarfrw}
R=12H^2 + 6\dot H\, .
\end{equation}
Since we are interested in the inflationary epoch, we shall assume
that this occurs when the slow-roll approximation holds true,
which is materialized by the following conditions,
\begin{equation}\label{slowrollconditionshubble}
\ddot{H}\ll H\dot{H},\,\,\, \frac{\dot{H}}{H^2}\ll 1\, ,
\end{equation}
and therefore, during this epoch, the Ricci scalar becomes
approximately,
\begin{equation}\label{ricciscalarapprox}
R\sim 12 H^2\, ,
\end{equation}
due to the fact that $\frac{\dot{H}}{H^2}\ll 1$. The inflationary
dynamical evolution is quantified in terms of the slow-roll
indices, $\epsilon_1$ ,$\epsilon_2$, $\epsilon_3$, $\epsilon_4$,
since the primordial curvature perturbations can be expressed in
terms of these. The slow-roll indices for $F(R)$ gravity can be
expressed as follows \cite{Hwang:2005hb,reviews1},
\begin{equation}
\label{restofparametersfr}\epsilon_1=-\frac{\dot{H}}{H^2}, \quad
\epsilon_2=0\, ,\quad \epsilon_3= \frac{\dot{F}_R}{2HF_R}\, ,\quad
\epsilon_4=\frac{\ddot{F}_R}{H\dot{F}_R}\,
 .
\end{equation}
Note that $\epsilon_2$ is in general defined as
$\epsilon_2=\frac{\ddot{\phi}}{\dot{\phi}H}$, so in the $F(R)$
gravity case it is equal to zero. When the slow-roll era is
materialized during the inflationary era, the slow-roll indices
satisfy the constraint $\epsilon_i\ll 1$, $i=1,3,4$ and the
primordial curvature perturbations are expressed as a perturbation
expansion with respect to the slow-roll indices. During the
slow-roll era, the observational indices of inflation, namely the
spectral index of scalar perturbations $n_s$ and the
tensor-to-scalar ratio $r$, can be expressed in terms of the
slow-roll indices as follows \cite{reviews1,Hwang:2005hb},
\begin{equation}
\label{epsilonall} n_s=
1-\frac{4\epsilon_1-2\epsilon_3+2\epsilon_4}{1-\epsilon_1},\quad
r=48\frac{\epsilon_3^2}{(1+\epsilon_3)^2}\, .
\end{equation}
Let us focus on the $F(R)$ gravity case, and let us start with the
tensor-to-scalar ratio, which is the ratio of the tensor
perturbations $P_T$ over the scalar perturbation $P_S$,
\begin{equation}\label{tensorananalytic}
r=\frac{P_T}{P_S}=8 \kappa^2 \frac{Q_s}{F_R}\, ,
\end{equation}
with,
\begin{equation}
\label{qsfrpreliminary}
Q_s=\frac{3\dot{F_R}^2}{2F_RH^2\kappa^2(1+\epsilon_3)^2}\, .
\end{equation}
Upon combining Eqs. (\ref{tensorananalytic}) and
(\ref{qsfrpreliminary}) we get,
\begin{equation}\label{ranalyticfinal}
r=48 \frac{\dot{F_R}^2}{4F_R^2H^2(1+\epsilon_3)^2}\, ,
\end{equation}
and due to the fact that $\epsilon_3= \frac{\dot{F}_R}{2HF_R}$, we
finally get,
\begin{equation}\label{ranalyticfinal1}
r=48\frac{\epsilon_3^2}{(1+\epsilon_3)^2}\, ,
\end{equation}
which is the expression for the tensor-to-scalar ratio given in
Eq. (\ref{epsilonall}). From the Raychaudhuri equation in the case
of a pure $F(R)$ gravity, we have,
\begin{equation}\label{approx1}
\epsilon_1=-\epsilon_3(1-\epsilon_4)\, ,
\end{equation}
and in the slow-roll approximation we have $\epsilon_1\simeq
-\epsilon_3$, hence the spectral index of the scalar perturbations
becomes,
\begin{equation}
\label{spectralfinal} n_s\simeq 1-6\epsilon_1-2\epsilon_4\, ,
\end{equation}
and the tensor-to-scalar ratio takes the form $r\simeq 48
\epsilon_3^2$, and due to the fact that $\epsilon_1\simeq
-\epsilon_3$, we finally have,
\begin{equation}
\label{tensorfinal} r\simeq 48\epsilon_1^2\, .
\end{equation}
The calculation of the slow-roll index $\epsilon_4$ is vital for
our analysis, so we focus on this now. Recall its functional form
is $\epsilon_4=\frac{\ddot{F}_R}{H\dot{F}_R}$ and as we will show,
it can be expressed in terms of the slow-roll index $\epsilon_1$.
We have,
\begin{equation}\label{epsilon41}
\epsilon_4=\frac{\ddot{F}_R}{H\dot{F}_R}=\frac{\frac{d}{d
t}\left(F_{RR}\dot{R}\right)}{HF_{RR}\dot{R}}=\frac{F_{RRR}\dot{R}^2+F_{RR}\frac{d
(\dot{R})}{d t}}{HF_{RR}\dot{R}}\, ,
\end{equation}
but $\dot{R}$ is,
\begin{equation}\label{rdot}
\dot{R}=24\dot{H}H+6\ddot{H}\simeq 24H\dot{H}=-24H^3\epsilon_1\, ,
\end{equation}
due to the fact that the slow-roll approximation $\ddot{H}\ll H
\dot{H}$ applies. Combining Eqs. (\ref{rdot}) and
(\ref{epsilon41}), after some algebra we get,
\begin{equation}\label{epsilon4final}
\epsilon_4\simeq -\frac{24
F_{RRR}H^2}{F_{RR}}\epsilon_1-3\epsilon_1+\frac{\dot{\epsilon}_1}{H\epsilon_1}\,
,
\end{equation}
however $\dot{\epsilon}_1$ is equal to,
\begin{equation}\label{epsilon1newfiles}
\dot{\epsilon}_1=-\frac{\ddot{H}H^2-2\dot{H}^2H}{H^4}=-\frac{\ddot{H}}{H^2}+\frac{2\dot{H}^2}{H^3}\simeq
2H \epsilon_1^2\, ,
\end{equation}
hence an approximation for the slow-roll index $\epsilon_4$ is,
\begin{equation}\label{finalapproxepsilon4}
\epsilon_4\simeq -\frac{24
F_{RRR}H^2}{F_{RR}}\epsilon_1-\epsilon_1\, .
\end{equation}
As it can be seen, $\epsilon_4$ can be expressed in terms of the
dimensionless parameter $x$, which is defined as follows,
\begin{equation}\label{parameterx}
x=\frac{48 F_{RRR}H^2}{F_{RR}}\, ,
\end{equation}
and in terms of $x$, the slow-roll $\epsilon_4$ is written as
follows,
\begin{equation}\label{epsilon4finalnew}
\epsilon_4\simeq -\frac{x}{2}\epsilon_1-\epsilon_1\, .
\end{equation}
At this point it is worth discussing the leading order
approximations we made. Below Eq. (\ref{approx1}) we made the
approximation $\epsilon_1\simeq -\epsilon_3$ which holds true at
leading order, and specifically at first order in the perturbative
expansion of the slow-roll index $\epsilon_1$ in terms of
$\epsilon_3$ and $\epsilon_4$, where the complete relation is
given in Eq. (\ref{approx1}). Thus the relation $\epsilon_1\simeq
-\epsilon_3$ holds true at first order in the slow-roll
perturbative expansion. Now the parameter $x$ itself contains
third derivatives $F_{RRR}$, and also derivatives $F_{RR}$ hence
can be in principle large. If we look at Eq. (\ref{epsilon4final})
we can see that $x$ naturally emerges in the expression of
$\epsilon_4$ (recall that $R\sim 12 H^2$ during the slow-roll
era), and the relation $\epsilon_4$ holds true so long as the
perturbative expansion still holds true for $\epsilon_4$, that is,
as long as $\epsilon_4$ is still smaller than unity. Thus upon
using the fact $\ddot{H}\ll H^2$ in Eq. (\ref{epsilon1newfiles})
we get Eq. (\ref{finalapproxepsilon4}) and expressed in terms of
$x$ one gets (\ref{epsilon4finalnew}). Note that no truncation in
terms of $x$ was made in order to get Eq.
(\ref{epsilon4finalnew}). The parameter $x$ naturally arises from
Eq. (\ref{epsilon4final}) if one omits terms that contain
$\ddot{H}$. Thus no truncation in terms of $x$ was made in order
to get Eq. (\ref{epsilon4finalnew}). But Eq.
(\ref{epsilon4finalnew}) holds true once the perturbation
expansion holds true and once the slow-roll conditions are
satisfied. For theories that yield $x$ too large, the slow-roll
expansion breaks and these theories cannot describe a viable
slow-roll inflationary era. Regarding the truncation of
$\epsilon_1$, namely, $\epsilon_1\simeq -\epsilon_3$, it holds at
first order in the expansion of the first slow-roll index in terms
of the slow-roll parameters. Note that if $x$ is too large, this
expansion breaks down, and thus we no longer have
$\epsilon_1\simeq -\epsilon_3$, but still we have Eq.
(\ref{approx1}) holding true, since this equation is derived from
the field equations, it is an exact relation. By combining Eqs.
(\ref{epsilon4finalnew}) and (\ref{spectralfinal}), the spectral
index of the primordial scalar curvature perturbations  takes the
final form,
\begin{equation}\label{asxeto1}
n_s-1=-4\epsilon_1+x\epsilon_1\, .
\end{equation}
Now, one can solve the above equation with respect to $\epsilon_1$
to obtain,
\begin{equation}\label{spectralasfunctionofepsilon1}
\epsilon_1=\frac{1-n_s}{4-x}\, ,
\end{equation}
and by substituting $\epsilon_1$ in the tensor-to-scalar ratio in
Eq. (\ref{tensorfinal}), we have,
\begin{equation}\label{mainequation}
r\simeq \frac{48 (1-n_s)^2}{(4-x)^2}\, .
\end{equation}
Now one can express the dimensionless parameter $x$ defined in Eq.
(\ref{parameterx}) in terms of the Ricci scalar and not the Hubble
rate during the slow-roll inflationary era, by making use of Eq.
(\ref{ricciscalarapprox}), so the parameter $x$ in terms of $R$ is
expressed as follows,
\begin{equation}\label{parameterxfinal}
x=\frac{4 F_{RRR}\,R}{F_{RR}}\, .
\end{equation}
In general, for inflationary dynamics purposes, one needs to
evaluate $x$ and $\epsilon_1$ at the first horizon crossing time
instance, and determine whether the inflationary dynamics is
viable by calculating $r$ and $n_s$ from Eqs. (\ref{mainequation})
and (\ref{asxeto1}) respectively. The parameter $x$ is not a
constant in general, and it can take various arbitrary values.
However in the next section we shall focus on the values it can
take in order for an $F(R)$ gravity to be considered a consistent
model and in order for a viable era to be produced.

Before closing this section, let us discuss the values and
behavior of the parameter $x$ defined in Eq.
(\ref{parameterxfinal}). As we mentioned, this parameter is not a
constant in general, it can be a constant though for only one
functional form of the function $F(R)$. This case will be analyzed
in section V and if $x$ is equal to a constant number $n$, that is
$x=-n$, by solving the differential equation $x=-n$ in terms of
the function $F(R)$, one gets,
\begin{equation}\label{constnatxolrevision}
F(R)=c_1+c_3 R+c_2\frac{16 c_1 R^{2-\frac{n}{4}}}{(n-8) (n-4)}\, ,
\end{equation}
with $c_3,c_2,c_1$ being integration constants. This is the only
case that $x$ can be a constant. In all other cases, the parameter
$x$ is in general a function of the Ricci scalar, thus $x=g(R)$
because it depends on the derivatives of $F(R)$ and the Ricci
scalar itself. In this work we shall be interested in various
asymptotic forms that $x$ might take, and we shall consider
analytic scenarios for which the differential equation $x=\frac{4
F_{RRR}\,R}{F_{RR}}=g(R)$ can be solved analytically.

\subsection{Viable $F(R)$ Gravity Inflation and Constraints on the $F(R)$ Gravity Form: The Exceptional Role of $R^2$ Gravity}

In this section we shall consider the allowed values of $x$ for
which the $F(R)$ gravity consistency relations are satisfied and
also we narrow down the allowed parameter space for the
dimensionless parameter $x$ in order to produce a viable
inflationary era. Before starting, let us first consider two cases
of interest, which are very simple to discuss. The first case is
the scenario for which $x$ is exactly equal to zero, in which case
by solving Eq. (\ref{parameterxfinal}) for $x=0$ we get that,
\begin{equation}\label{r2inflation}
F(R)=R+c_1\,R^2=R+\frac{R^2}{M^2}\, ,
\end{equation}
where $c_1$ is an integration constant, which can be chosen to be
$c_1=\frac{1}{M^2}$ due to the relevance of the $R^2$ gravity with
inflation. As it proves, $M$ is determined by the amplitude of the
scalar perturbations, as we discuss later on in this section and
it is basically an integration constant, and not a fundamental
mass scale like the cosmological constant.

Another value of interest is when the tensor-to-scalar ratio
(\ref{mainequation}) blows up, which occurs for $x=4$. This case,
and in general the case with $x=\mathrm{const}$ is problematic,
because if we solve (\ref{parameterxfinal}) to be a constant,
namely $x=n$, we get the general solution,
\begin{equation}\label{generalsoluionpowerlaw}
F(R)=c_2+c_3 R+\frac{16 c_1 R^{2-\frac{n}{4}}}{(n-8) (n-4)}\, ,
\end{equation}
with $c_i$, $i=1,2,3$ being integration constants. This case,
along with power-law $F(R)$ gravity models, will be dealt in a
later section, separately, since there are important issues to
discuss about it.

Hence we need to clarify the meaning that $x$ approaches a
specific value asymptotically but cannot be exactly equal to a
constant. This means that in general $x$ can take the form,
\begin{equation}\label{asymptoticx}
x\sim n \beta(R)\, ,
\end{equation}
and asymptotically, for large curvatures, the function $\beta (R)$
may approach zero, or unity or some other allowed constant. For
example, the value $x=4$ may be approximated by
$x=4\,\left(\frac{R}{\Lambda}\right)^{\epsilon}$ with
$\frac{R}{\Lambda}$ being $\frac{R}{\Lambda}\gg 1$ and
$\epsilon\ll 1$, in the large curvature limit, and $\Lambda$ is
the cosmological constant. In this case, no simple power-law
gravity can generate the $x\sim 4$ case, and we will show later on
some scenarios of this sort. In this case, the $x\sim 4$ scenario
describes a scale invariant power spectrum as it can be seen from
Eq. (\ref{asxeto1}). But this is a peculiar situation in which one
cannot use the relation (\ref{spectralasfunctionofepsilon1}) which
diverges. This case must be dealt separately.

There is a caveat however, in the case $x\sim 4$, since as we now
show, $x$ is not allowed to take such values, if one requires a
consistent $F(R)$ gravity. Let us show this in detail, and we also
determine the values of $x$ for which one may obtain a
self-consistent $F(R)$ gravity description. The values of $x$ are
constrained by the de Sitter stability criterion
(\ref{criterion3}). If one requires that the scalaron mass is
always $m^2\geq 0$ in Eq. (\ref{scalaronmassfinal}), then the
criterion (\ref{criterion3}) must hold true. In order to ensure
$m^2\geq 0$, and also to ensure that the scalaron mass decreases
as the curvature decreases and the conversely, the scalaron mass
increases as the curvature increases in the large curvature
regime, one must require that the de Sitter scalaron mass, is
monotonically increasing, or zero, in order to cover also the
extremum case. Thus the derivative of $m^2(R)$, must satisfy,
\begin{equation}\label{derivativescalaronmass}
\frac{\partial m^2(R)}{\partial R}\geq 0 \, .
\end{equation}
Remarkably, as we will see, this requirement also affects the late
time behavior of the models. Let us analyze in brief the
requirement (\ref{derivativescalaronmass}) as it proves to be of
fundamental importance. What we basically require with the
condition (\ref{derivativescalaronmass}) is that the de Sitter
scalaron mass $m^2(R)$ is monotonically increasing, or has an
extremum of global type. Remarkably, as it also proves, for the
viable models that satisfy this constraint, the de Sitter scalaron
mass is large at high curvatures and small at low curvatures,
which is important if someone needs the $F(R)$ gravity to describe
both late and early-time de Sitter evolutions. Then the decreasing
scalaron mass with decreasing curvature indicates that the scalar
degree of freedom becomes lighter and can mediate interactions
over longer distances. This is why $F(R)$ models are effective at
explaining phenomena such as the accelerated expansion of the
Universe at low curvatures. Let us note that this is the first
time in the literature that the requirement
(\ref{derivativescalaronmass}) is imposed on potential $F(R)$
gravities. Let us evaluate $\frac{\partial m^2(R)}{\partial R}$,
so we have,
\begin{equation}\label{equationderivativescalaron}
\frac{\partial m^2(R)}{\partial
R}=-\frac{1}{12}\frac{F_R}{R\,F_{RR}}\,\frac{4\,R\,F_{RRR}}{F_{RR}}
\, ,
\end{equation}
which can be expressed in terms of the parameters $y$ and $x$
defined in Eqs. (\ref{yparameterdefinition}) and
(\ref{parameterxfinal}) respectively, as follows,
\begin{equation}\label{equationderivativescalaronfinal}
\frac{\partial  m^2(R)}{\partial R}=-\frac{x}{3y} \, .
\end{equation}
If one requires the condition (\ref{derivativescalaronmass}),
simultaneously with the de Sitter stability condition
(\ref{criterion3}), one gets the following constraints for the
values of $x$, depending on the values of $y$,
\begin{equation}\label{constraintx2}
x\leq 0,\,\,\,0< y\leq 1
\end{equation}
Thus we have one condition for $x$, it must be either zero or a
negative number. At this point let us further elaborate on the
requirement (\ref{derivativescalaronmass}), which may seen as a
non-naturally imposed condition that may falsify the model
agnostic approach in $F(R)$ gravity inflation we try to introduce
in this work. In a nutshell, the monotonicity criterion stems from
stability requirements, screening requirements and desirable
long-range features of the dark energy era, as we now evince in
detail. The assumption that the scalaron mass \( m^2(R) \) must be
a monotonically increasing function of curvature is theoretically
motivated for the following reasons, firstly, it guarantees the
stability of both the early- and late-time de Sitter phases,
secondly, it proves that it naturally arises in all inflationary
viable models that also unify inflation with dark energy. Thirdly,
it ensures a consistent screening mechanism via chameleon
suppression at high curvature and lastly, it allows the scalaron
to become light at late times and thus dynamically mimic dark
energy. Regarding the issue on how the constraint
(\ref{derivativescalaronmass}) affects the overall model agnostic
approach we try to achieve in this work, we need to mention that
the requirement (\ref{derivativescalaronmass}) indeed filters the
landscape of possible models, but it retains full generality
within the space of physically meaningful and observationally
consistent theories, preserving the model-agnostic spirit of the
formalism. We came to this conclusion by observing in the
literature the behavior of the viable models of $F(R)$ gravity
inflation that can also describe successfully a late time era,
thus unifying inflation with the dark energy era. Thus the
condition (\ref{derivativescalaronmass}) is theoretically
motivated by requiring the late-time scalaron mass to be smaller
than the early-time scalaron mass, which again we found that it
holds true in viable unification models. Let us expand the above
discussion with more details. The assumption that \( m^2(R) \) is
monotonically increasing with the Ricci scalar \( R \) is not
arbitrary from a theoretical perspective, but it is rather
physically and phenomenologically motivated. It ensures dynamical
mass suppression of the scalaron at low curvatures (late times),
thus allowing it to mediate large-scale gravitational effects such
as dark energy. In the  high curvature regime (early times), a
large scalaron mass, stabilizes the de Sitter inflationary vacuum,
and thus suppresses unwanted scalar fluctuations. This said
behavior is met to all known viable \( F(R) \) models that aim to
unify inflation and late-time acceleration, as we observed in the
models available in the literature. The monotonicity condition
does not fully constrain the theory, but it rather acts as a
filter, choosing viable classes of models. Also, the monotonicity
criterion supports stability to screening mechanisms, such as the
chameleon effect. Indeed, viable \( F(R) \) dark energy theories
must pass solar system tests, which require that the scalaron must
be heavy in high-density environments (i.e., where the Ricci
scalar \( R \) is large). A monotonically increasing scalaron mass
\( m^2(R) \) ensures that the scalaron is naturally screened,
which is the most salient feature of the chameleon mechanism.
Conversely, non-monotonicity may lead to unscreened scalar field
propagation in high-curvature environments, thus violating local
gravity empirical constraints.

But there is also more strong theoretical motivation for the
monotonicity criterion, having to do again with the unification of
inflation and dark energy era. In an $F(R)$ gravity model that is
required to describe successfully both the inflationary era and
the dark energy era, the scalaron mass should have this
monotonicity property. Specifically, the scalaron mass determines
the Compton wavelength \( \lambda_c \sim m^{-1} \) of the scalar
degree of freedom. At large \( m^2(R) \), the scalaron is heavy
and its influence is short-range. At small \( m^2(R) \), it
mediates long-range interactions and can drive cosmic
acceleration. As we will show, this decreasing trend of \( m^2(R)
\) with decreasing \( R \) is universal among all viable
unification models. Another theoretical motivation comes from the
fact that a monotonically decreasing scalaron mass, ensures that
tachyonic instabilities do not occur at early times. To ensure
stability of de Sitter vacua and suppression of tachyonic
instabilities, one must require:
\[
m^2(R) \geq 0
\]
and furthermore,
\[
\frac{\partial m^2}{\partial R} \geq 0
\]
This ensures that high-curvature regimes (during inflation) are
stable attractors. If the scalaron mass \( m^2(R) \) decreased
with the curvature, one could encounter tachyonic instabilities or
even strong infrared modifications at early times-both physically
problematic. Hence, while the monotonicity condition constrains
the full space of available \( F(R) \) functions, it does not
render the formalism model-specific. It merely acts as a
physically motivated selection criterion-similar to the slow-roll
conditions in scalar field inflation. Thus our approach remains
agnostic within the domain of phenomenologically viable and
theoretically consistent models, and hence preserves to some
extent the ``model-agnostic'' character. Thus, the parameters $x$
and $y$ must be equal to some appropriate forms of the following
type,
\begin{equation}\label{typei}
x\sim -n\beta_1(R,\Lambda)\, ,
\end{equation}
and
\begin{equation}\label{typeii}
y\sim -n\beta_2(R,\Lambda)\, ,
\end{equation}
with the functions $\beta_1(R,\Lambda)$ and $\beta_2(R,\Lambda)$
being appropriate functions. Now recalling the functional form of
the spectral index, namely Eq. (\ref{asxeto1}), and assuming that
sensible models of inflationary $F(R)$ gravity will yield a first
slow-roll index of the order $\epsilon_1\sim
\mathcal{O}(10^{-3})$, it makes sense that $x$ will be in the
range,
\begin{equation}\label{extraconstraint2025}
-1\leq x\leq 0\, .
\end{equation}
We shall further discuss this issue later on. Hence, taking this
into account, and also that $0<y\leq 1$, the functions
$\beta_i(R,\Lambda)$ will yield value $0<\beta_i(R,\Lambda)<1$
when evaluated at the first horizon crossing and also $0<n<1$.

Let us now consider possible forms of the general function
$\beta_1(R,\Lambda)$, so one may consider simple positive
functions for which the differential equation
$x=-n\beta_1(R,\Lambda)$ can be solved analytically. Thus a
general form for the parameter $x$ can be the following,
\begin{equation}\label{generalformofx}
x=-n \left(\frac{R}{\Lambda} \right)^{\epsilon}\, .
\end{equation}
Other forms for the function $\beta_1(R,\Lambda)$ can be
exponentials, but this case cannot be solved analytically. During
the slow-roll era, the fraction $R/\Lambda$ is of the order
$R/\Lambda \sim 10^{111}$, thus there are two asymptotic scenarios
of interest. One that $\epsilon<0$, in which case $\lim_{R\to
\infty} \beta_1(R,\Lambda)=\lim_{R\to \infty}
\left(\frac{R}{\Lambda} \right)^{-|\epsilon|}\sim 0$, which is
compatible with the constraint (\ref{constraintx2}), and the other
asymptotic case is when $\epsilon>0$, in which case only when
$\epsilon \ll 1$, one may obtain a value for $x$ which is
compatible with the constraints (\ref{constraintx2}) and
(\ref{extraconstraint2025}). Thus when $\epsilon \ll 1$, the
approximate value of $x$ is $x\sim -n$. This result is of great
importance, since these two cases are basically the attractors of
any viable $F(R)$ gravity inflation, and basically correspond to
the Starobinsky inflation and $\alpha$-attractor potentials in the
Einstein frame. In the case of the $R^2$ attractor solution, any
scenario which will lead to a value of $x\ll 1$ at first horizon
crossing, this scenario will yield an inflationary evolution
identical to the Starobinsky inflation. So the Starobinsky
inflation is an attractor of $F(R)$ gravity inflation, and this
occurs for any $F(R)$ gravity that yields $x\sim R^{-\epsilon}$,
$\epsilon>0$. In this case, one has,
\begin{equation}\label{rnsstarobinsky}
r\sim 3 (1-n_s)^2\, ,
\end{equation}
which describes  the $r-n_s$ relation obeyed by the Starobinsky
inflation model, which corresponds to the case $x=0$. Now if $x$
is somewhere in the range $-1<x<0$, then one gets an
$\alpha$-attractor like behavior, since $x$ is basically negative,
and in effect $r$ can be smaller than in the Starobinsky scenario.
In this case, the $r-n_s$ relation takes the form,
\begin{equation}\label{alphaattractor}
r\sim 3\alpha (1-n_s)^2\, ,
\end{equation}
with $\alpha=\frac{16}{(4-x)^2}$, which is identical to the
$\alpha$-attractors relation
\cite{alpha1,alpha2,alpha3,alpha4,alpha5,alpha6,alpha7,alpha8,alpha9,alpha10,alpha10a,alpha11,alpha12,vernov}.
Note however that in order to have a viable inflationary theory,
the first slow-roll index must be smaller than unity at first
horizon crossing, so not all theories that yield a value for the
parameter $x$ in the range  $-1<x<0$, yield a viable quasi-de
Sitter solution. Caution is thus needed in this respect.

This is a somewhat important issue, since there maybe exist
theories that may yield a small $x$, nearly zero, or even in the
range  $-1<x<0$, but it is not certain that the first slow-roll
index, and therefore the spectral index may ,be observationally
acceptable. One such example is the exponential case we shall
briefly discuss in the next section. In most scenarios in which
$x\sim -R^{\epsilon}$, the $F(R)$ gravities that produce such
values are deformations of $R^2$ gravity, so a quasi-de Sitter
inflation theory is produced. It is worth recalling the essential
features of the $R^2$-inflation dynamical evolution. This will
prove to be valuable in order to have an idea of how large can the
first slow-roll inflationary index be in the context of $R^2$
inflation deformations. For the $R^2$ gravity, of the form,
\begin{equation}\label{effectivelagrangian2}
F(R)=R+\frac{1}{M^2}R^2\, ,
\end{equation}
the Friedmann equation reads,
\begin{equation}\label{diffeqndefomedstaro}
\ddot{H}+3 H \dot{H}-\frac{\dot{H}^2}{2 H}+\frac{1}{12} M^2 H=0\,
,
\end{equation}
and due to the slow-roll approximation, one has,
\begin{equation}\label{finalformfriedmanneqn}
\dot{H}\simeq -\frac{1}{36} (2) M^2\, ,
\end{equation}
which can easily be solved,
\begin{equation}\label{quasidesitter}
H(t)=H_I-\frac{1}{36} t \left(2M^2\right)\, ,
\end{equation}
with $H_I$ being an integration constant, and actually is the
inflationary scale. The evolution (\ref{quasidesitter}) is a
quasi-de Sitter evolution. Now, for the above quasi-de Sitter
evolution one has,
\begin{equation}\label{epsilon1indexanalytic}
\epsilon_1=-\frac{-2M^2}{36 \left(H_I-\frac{1}{36} t \left(
2M^2\right)\right)^2}\, ,
\end{equation}
and we can readily find the time instances that inflation starts
and ends, $t_i$ and $t_f$ respectively. By solving
$\epsilon_1(t_f)=1$, we get,
\begin{equation}\label{finaltimeinstance}
t_f=\frac{6 \left(6 H_I  M^2+6 H_I M^2-\sqrt{ ^3 M^6+3 M^6+3
 M^6+M^6}\right)}{ M^4+2 M^4+M^4}\, ,
\end{equation}
and since the $e$-foldings number $N$ is,
\begin{equation}\label{efoldingsnumber}
N=\int_{t_i}^{t_f}H(t)dt\, ,
\end{equation}
from it we obtain $t_i$ for the quasi-de Sitter evolution
(\ref{quasidesitter}),
\begin{equation}\label{ti}
t_i=\frac{6 \left(6 H_I+\sqrt{(2) M^2 (2 N+1)}\right)}{(2) M^2}\,
.
\end{equation}
Hence, the first slow-roll index can be expressed in terms of the
$e$-foldings number,
\begin{equation}\label{epsilon1lambdaind}
\epsilon_1(t_i)=\frac{1}{2N}\, ,
\end{equation}
and the observational indices of inflation for the Starobinsky
model read $n_s\sim 1-\frac{2}{N}$ and $r\sim \frac{12}{N^2}$.
Now, for $N\sim 60$, one has $\epsilon_1\sim 0.0083$ and this is
compatible with the Planck 2018 constraints on the first slow-roll
index, where it is expected that  $\epsilon_1\sim
\mathcal{O}(10^{-3})$ \cite{Planck:2018jri}, however, the Planck
constraints are based on a single scalar field theory. Notably
though, the same constraint should apply for the Jordan frame
counterparts of scalar field theories, thus we expect that
$\epsilon_1\sim \mathcal{O}(10^{-3})$ \cite{Planck:2018jri} for
the viable $F(R)$ gravities, hence the constraint
(\ref{extraconstraint2025}) is justified according to this line of
reasoning. At this point, let us investigate in a model
independent way the effect of the parameter $x$ on the
inflationary indices and also we compare the results with Planck
data. To this end, we shall fix the first slow-roll index to have
three distinct values, namely $\epsilon_1=0.01$,
$\epsilon_1=0.001$ and $\epsilon_1=0.008$ and we analyze the
inflationary phenomenology in terms of values of $x$ in the range
$x=[-0.9,-0.1]$ using the relations (\ref{asxeto1}) and
(\ref{mainequation}). The Planck 2018 constraints on the scalar
spectral index and the tensor-to-scalar ratio are
\cite{Planck:2018jri},
\begin{equation}\label{observationaldatanewresults}
n_s = 0.9649 \pm 0.0042 \quad (68\%\ \text{CL}),\,\,\,r<0.064\, ,
\end{equation}
and now we will perform some confrontation with the Planck 2018
likelihood curves.
\begin{figure}[h!]
\centering
\includegraphics[width=18pc]{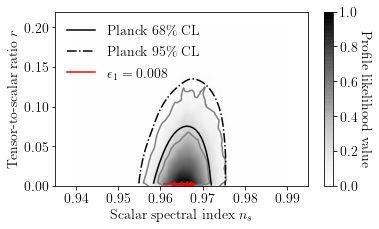}
\includegraphics[width=18pc]{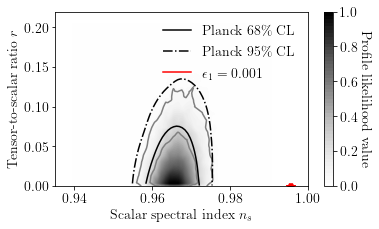}
\includegraphics[width=18pc]{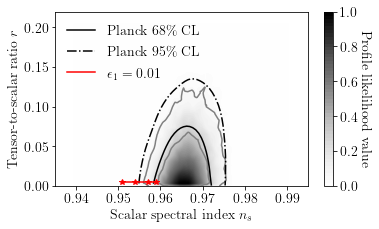}
\caption{The Planck 2018 likelihood curves, versus the $F(R)$
gravity phenomenology for three distinct values of the slow-roll
index $\epsilon_1$ and with $x$ in the range $x=[-0.9,-0.1]$. The
upper left plot corresponds to $\epsilon_1=0.008$, the upper right
to $\epsilon_1=0.001$ and the bottom plot to $\epsilon_1=0.01$.}
\label{plot1}
\end{figure}
Our results can be found in Fig. \ref{plot1}, where we present the
Planck 2018 likelihood curves, versus the $F(R)$ gravity
phenomenology for three distinct values of the slow-roll index
$\epsilon_1$, namely $\epsilon_1=0.01$, $\epsilon_1=0.001$ and
$\epsilon_1=0.008$ and with the parameter $x$ chosen in the range
$x=[-0.9,-0.1]$, using Eqs. (\ref{asxeto1}) and
(\ref{mainequation}). The upper left plot corresponds to
$\epsilon_1=0.008$, the upper right to $\epsilon_1=0.001$ and the
bottom plot to $\epsilon_1=0.01$. As it can be seen, values of
$\epsilon_1$ near the ones obtained for the Starobinsky model are
optimal and produce the most viable inflationary phenomenology for
$x$ in the range $x=[-0.9,-0.1]$ which is compatible with the
constraint (\ref{extraconstraint2025}).

\section{Viable Inflation in $F(R)$ Gravity}

In this section we shall analyze all the possible scenarios that
can yield a viable $F(R)$ gravity inflationary era. We will focus
on solutions for which the differential equation
$x=-n\beta(R,\Lambda)$ can be solved analytically.

\subsection{Models with $x=-n\,\left(\frac{R}{\Lambda}\right)^{-m}$, with $m>0$ and the Exceptional Role of $R^2$ Gravity}

Our first analysis will involve cases in which
$x=-n\,\left(\frac{R}{\Lambda}\right)^{-m}$ with $m$ being some
positive number, integer or non-integer. As we shall see, this
case enables us to evaluate analytically the form of $F(R)$
gravity which yields a parameter $x$ to be of the form
$x=-n\,\left(\frac{R}{\Lambda}\right)^{-m}$. In this class of
models belong models which can yield asymptotically
$x=-n\,\left(\frac{R}{\Lambda}\right)^{-m}$ in the large curvature
limit, which applies during the inflationary era. We shall study
several cases, in which $m$ can be an integer, or some fraction.

\subsubsection{Models with $x=-n\,\left(\frac{R}{\Lambda}\right)^{-m}$ and $m$ is a Positive Integer}

Let us consider the case in which $m$ is some positive integer.
The general case cannot be solved analytically, so we will examine
some characteristic cases, with $m=1,2,3,4,5$ which yield some
results in closed form.

Let us first consider the case with $m=1$, so by solving
$x=-n\,\left(\frac{R}{\Lambda}\right)^{-1}$, we obtain
analytically the following solution,
\begin{equation}\label{sol1}
F(R)=c_3 R+c_2+\frac{1}{32} c_1 \left(\Lambda  n (\Lambda  n-8 R)
\, \text{Ei}\left(\frac{n \Lambda }{4 R}\right)+4 R
e^{\frac{\Lambda n}{4 R}} (4 R-\Lambda  n)\right)\, ,
\end{equation}
where $c_i$, $i=1,2,3$ are integration constants, and the function
$\text{Ei}(z)$ is the exponential integral. Note that $c_2$ plays
the role of a cosmological constant, $c_3$ is a multiplication
fact of the standard Einstein-Hilbert term, a rescaled
Einstein-Hilbert term, and $c_1$ affects the amplitude of the
scalar perturbations. This behavior of the integration constant
applies in all the functional forms of $F(R)$ we shall present in
the rest of the article. Clearly this form of $F(R)$ gravity
contains Einstein-Hilbert gravity, with $c_3=1$, or some rescaled
form of Einstein-Hilbert gravity with $c_3\neq 1$. More
importantly, the $F(R)$ gravity of Eq. (\ref{sol1}) is basically a
deformation of $R^2$ gravity during the inflationary era. This is
not difficult at all to imagine, since $R\sim 10^{44}\,$eV$^2$
(Taking the inflationary scale to be $H_I\sim 10^{16}\,$GeV)
during inflation, and also the cosmological constant is of the
order $\Lambda\sim 10^{-67}$eV$^2$ thus during inflation, the
fraction $\frac{R}{\Lambda}$ is of the order,
\begin{equation}\label{roverlambda}
\frac{R}{\Lambda}\sim 10^{111}\, ,
\end{equation}
so it is basically huge. Thus $\frac{\Lambda}{R}$ is basically
zero and thus the exponential becomes nearly $e^{\frac{\Lambda
n}{4 R}}\sim 1$, therefore, the $F(R)$ gravity (\ref{sol1}) during
inflation asymptotically becomes,
\begin{equation}\label{asymsol1}
F(R)\sim R+c_2 -\frac{1}{8} c_1 \Lambda  n R+\frac{c_1 R^2}{2}\, ,
\end{equation}
so by keeping the dominant terms ($-\frac{1}{8} c_1 \Lambda  n
R\ll \frac{c_1 R^2}{2}$), the dominant $F(R)$ gravity is an $R^2$
gravity,
\begin{equation}\label{asymsol12}
F(R)\sim  R+c_2 +\frac{R^2}{M^2}\, .
\end{equation}
where we set $c_1=\frac{2}{M^2}$. The parameter $M$ can be
constrained by the amplitude of the primordial scalar
perturbations, so we will find its value later on. The important
issue to note is that the $F(R)$ gravity function (\ref{sol1}) is
nothing but an $R^2$ gravity during inflation. It is also
important to note that although the exponential and the
exponential integral functions are subdominant primordially, these
might become important at late times. In fact as we show in a
later section, this is exactly the case, so our formalism provides
formally an $F(R)$ gravity that can describe simultaneously
inflation and the dark energy era. As a final comment for this
model, let us see if the de Sitter constraints are satisfied.
Indeed, since $\frac{\Lambda}{R}\ll 1$ during inflation, this
model yields $x\sim 0$ and also for this model during inflation we
have $y\sim 1$, so the constraint (\ref{constraintx2}) is
satisfied. In fact, in this case, the de Sitter mass has an
extremum and also the de Sitter mass is nearly equal to zero,
which means that the Einstein frame potential is nearly flat. This
is exactly what happens for $R^2$ gravity. One important feature
to note is that the scalaron mass for the model under
consideration is small at late times and large at early times.
This is compatible with the requirement that the same $F(R)$
gravity theory should describe late times and early times. This
behavior is reported for the first time in the $F(R)$ gravity
literature.

A similar model of this sort is the following,
\begin{equation}\label{modelforde}
F(R)= R+\frac{R^2}{M^2}+\lambda  R \exp \left(\frac{\Lambda
\epsilon }{R}\right)-\frac{\Lambda
\left(\frac{R}{m_s^2}\right)^{\delta }}{\gamma }+\lambda \Lambda
\epsilon\, ,
\end{equation}
with $\gamma$, $\delta$ and $\epsilon$ being dimensionless
parameters, and $0<\delta<1$.  Note that the parameter $M$ affects
the amplitude of the scalar perturbations. The parameter $x$ for
the model (\ref{modelforde}) is equal to,
\begin{equation}\label{solex1}
x=\frac{4 \Lambda  M^2 \left(\gamma  \lambda  \Lambda  \epsilon ^2
(\Lambda  \epsilon -3 R)-\delta  \left(\delta ^2-3 \delta
+2\right) R^2 \left(\frac{R}{m_s^2}\right)^{\delta }
e^{\frac{\Lambda  \epsilon }{R}}\right)}{R \left(\gamma  \lambda
\Lambda ^2 M^2 \epsilon ^2+R e^{\frac{\Lambda  \epsilon }{R}}
\left(2 \gamma  R^2-(\delta -1) \delta  \Lambda  M^2
\left(\frac{R}{m_s^2}\right)^{\delta }\right)\right)}\, ,
\end{equation}
which during inflation, and thus in the large curvature limit,
becomes asymptotically,
\begin{equation}\label{xextrasol1}
x\simeq -\frac{2 \delta  \left(\delta ^2-3 \delta +2\right)
\Lambda  M^2 m_s^{-2\delta } R^{\delta -2}}{\gamma }\, ,
\end{equation}
which also is nearly equal to zero, namely $x\sim 0$ and also $y$
for the model (\ref{modelforde}) is,
\begin{equation}\label{yforextramodel}
y=\frac{\gamma  \lambda  \Lambda ^2 M^2 \epsilon ^2+R
e^{\frac{\Lambda  \epsilon }{R}} \left(2 \gamma  R^2-(\delta -1)
\delta  \Lambda  M^2 \left(\frac{R}{m_s^2}\right)^{\delta
}\right)}{R \left(e^{\frac{\Lambda  \epsilon }{R}} \left(M^2
\left(\gamma R-\delta  \Lambda
\left(\frac{R}{m_s^2}\right)^{\delta }\right)+2 \gamma
R^2\right)+\gamma  \lambda  M^2 (R+\Lambda \epsilon )\right)}\, ,
\end{equation}
which in the large curvature regime becomes $y\sim 1$, thus the
constraint of Eq. (\ref{constraintx2}) is satisfied. The model of
Eq. (\ref{modelforde}) is basically an $R^2$ model during
inflation, but it is great phenomenological importance, since the
subdominant terms during the inflationary era, become dominant at
late times and drive the evolution generating a successful dark
energy era. We shall demonstrate this in a later section. Note the
presence of the exponentials in both the model (\ref{sol1}) and
(\ref{modelforde}), and these are formally introduced since they
lead to a $x$ containing inverse powers of the curvature. Similar
models were used in Refs. \cite{Oikonomou:2020qah}, based on
phenomenological reasoning. In this article, the one of the major
breakthroughs is that models containing exponentials, like
(\ref{sol1}) and (\ref{modelforde}), formally emerge as
deformations of the $R^2$ model, which yield a parameter
$x=-n\,\left(\frac{R}{\Lambda}\right)^{-m}$.

As it proves, other values of $m$ in the parameter
$x=-n\,\left(\frac{R}{\Lambda}\right)^{-m}$ lead to $R^2$
inflation deformations. Let us give some characteristic examples
here. For $m=2$, by solving
$x=-n\,\left(\frac{R}{\Lambda}\right)^{-2}$, we obtain
analytically the following solution,
\begin{equation}\label{sol2}
F(R)=\frac{1}{16} c_1 \left(-4 \sqrt{2 \pi } \sqrt{n} R \Lambda \,
\text{erfi}\left(\frac{\sqrt{n} \Lambda}{2 \sqrt{2} R}\right)+n
\Lambda^2 \, \text{Ei}\left(\frac{n \Lambda^2}{8 R^2}\right)+8 R^2
e^{\frac{n \Lambda^2}{8 R^2}}\right)+c_3 R+c_2\, ,
\end{equation}
where again $c_i$, $i=1,2,3$ are integration constants, and the
functions $\text{Ei}(z)$ and $\text{erfi}(z)$ are the exponential
integral and the error function. Clearly this form of $F(R)$
gravity contains Einstein-Hilbert gravity too. In this case too,
the $F(R)$ gravity of Eq. (\ref{sol2}) is basically a deformation
of $R^2$ gravity during the inflationary era. Also in this case
too, the de Sitter constraints are satisfied, since
$\frac{\Lambda}{R}\ll 1$ during inflation, this model yields
$x\sim 0$ and also $y\sim 1$, so the constraint
(\ref{constraintx2}) is satisfied for this case too. Now for
$m=3,4,5...$ a problem occurs, since the equation
$x=-n\,\left(\frac{R}{\Lambda}\right)^{-m}$ for $m=3,4,5...$ leads
to complex functional forms. So for $m=3,4,5..$ and so on, we
shall solve the equation
$x=n\,\left(\frac{R}{\Lambda}\right)^{-m}$. This causes no
inconsistency because during inflation $x\sim 0$, however it is
notable that the de Sitter criterion (\ref{constraintx2}) will be
violated. So these models are peculiar since the scalaron mass is
not a monotonic function of the Ricci scalar which means that at
late-times one might have a problem describing the Universe in a
consistent way with these $F(R)$ gravities. Let us quote the
functional form of these $F(R)$ gravities, for $m=3,4,5$, which
again we note that these violate the criterion
(\ref{constraintx2}), so basically these are deemed non viable
phenomenologically.
\begin{itemize}
    \item For $m=3$ by
solving $x=n\,\left(\frac{R}{\Lambda}\right)^{-3}$, we obtain
analytically the following solution,
\begin{equation}\label{sol3}
F(R)=c_3 R+c_2+\frac{1}{2} c_1 R^2 e^{-\frac{\Lambda ^3 n}{12
R^3}}-\frac{c_1 R^2 \sqrt[3]{\frac{\Lambda ^3 n}{R^3}} \Gamma
\left(\frac{2}{3},\frac{n \Lambda ^3}{12 R^3}\right)}{2^{2/3}
\sqrt[3]{3}}+\frac{c_1 R^2 \left(\frac{\Lambda ^3
n}{R^3}\right)^{2/3} \Gamma \left(\frac{1}{3},\frac{n \Lambda
^3}{12 R^3}\right)}{4 \sqrt[3]{2} 3^{2/3}}\, ,
\end{equation}
where again $c_i$, $i=1,2,3$ are integration constants, and the
function $\Gamma(z,b)$ is the Gamma function. Clearly this
function is an $R^2$ deformation during inflation, basically an
$R^2$ gravity, but at late times the exponential functions are
subdominant, and so are the Gamma functions, so this model cannot
describe successfully a late-time evolution. This is what we
expected, since the criterion (\ref{constraintx2}) is violated for
this model, so at late times the scalaron mass has an undesired
behavior.
    \item For $m=4$ by
solving $x=n\,\left(\frac{R}{\Lambda}\right)^{-4}$, we obtain
analytically the following solution,
\begin{equation}\label{sol4}
F(R)=c_3 R+c_2+\frac{1}{8} c_1 \left(4 R^2 \left(e^{-\frac{\Lambda
^4 n}{16 R^4}}-\sqrt[4]{\frac{\Lambda ^4 n}{R^4}} \Gamma
\left(\frac{3}{4},\frac{n \Lambda ^4}{16
R^4}\right)\right)-\sqrt{\pi } \Lambda ^2 \sqrt{n} \,
\text{erf}\left(\frac{\Lambda ^2 \sqrt{n}}{4 R^2}\right)\right)\,
,
\end{equation}
where again $c_i$, $i=1,2,3$ are integration constants. Also $c_2$
plays the role of a cosmological constant, $c_3$ is a
multiplication fact of the standard Einstein-Hilbert term, and
$c_1$ affects the amplitude of the scalar perturbations. Clearly
this function is an $R^2$ deformation during inflation too,
basically an $R^2$ gravity, but at late times the exponential
functions are subdominant, and so are the Gamma functions and
error functions, so this model cannot describe successfully a
late-time evolution. As in the $m=3$ case, this is what we
expected because the criterion (\ref{constraintx2}) is violated in
this case too.
    \item For $m=5$ by
solving $x=n\,\left(\frac{R}{\Lambda}\right)^{-5}$, we obtain
analytically the following solution,
\begin{equation}\label{sol5}
F(R)=c_3 R+c_2+\frac{1}{2} c_1 R^2 e^{-\frac{\Lambda ^5 n}{20
R^5}}-\frac{c_1 R^2 \sqrt[5]{\frac{\Lambda ^5 n}{R^5}} \Gamma
\left(\frac{4}{5},\frac{n \Lambda ^5}{20 R^5}\right)}{2^{2/5}
\sqrt[5]{5}}+\frac{c_1 R^2 \left(\frac{\Lambda ^5
n}{R^5}\right)^{2/5} \Gamma \left(\frac{3}{5},\frac{n \Lambda
^5}{20 R^5}\right)}{2\ 2^{4/5} 5^{2/5}}\, ,
\end{equation}
where again $c_i$, $i=1,2,3$ are integration constants. This case
also shares the same characteristics as the $m=3,4$ cases quoted
above. There is some sort of regularity for the solutions as $m$
increases, to take values in the integers, which is notable.
\end{itemize}
Now let us consider scenarios in which $m$ is a rational number of
the form $m=\frac{k}{\alpha}$, with $k$ and $\alpha$ some positive
integers. In the case $k<\alpha$, the models that one obtains by
solving the equation $x=-n\,\left(\frac{R}{\Lambda}\right)^{-m}$
result to $R^2$ deformations during inflation which are also
consistent with the criterion (\ref{constraintx2}). Let us quote
here the cases $m=1/2$ and $m=1/3$,
\begin{itemize}
    \item For $m=1/2$, the solution of the equation
    $x=-n\,\left(\frac{R}{\Lambda}\right)^{-1/2}$ is,
\begin{equation}\label{sol512}
F(R)=c_3 R+c_2+\frac{1}{192} c_1 \left(\Lambda  n^2 \left(\Lambda
n^2-48 R\right) \, \text{Ei}\left(\frac{n \sqrt{\Lambda }}{2
\sqrt{R}}\right)+2 \sqrt{R} e^{\frac{\sqrt{\Lambda } n}{2
\sqrt{R}}} \left(-\Lambda ^{3/2} n^3-2 \Lambda  n^2 \sqrt{R}+40
\sqrt{\Lambda } n R+48 R^{3/2}\right)\right)\, ,
\end{equation}
    \item For $m=1/3$, the solution of the equation
    $x=-n\,\left(\frac{R}{\Lambda}\right)^{-1/3}$ is,
\begin{align}\label{sol5121}
& F(R)=c_3 R+c_2+\frac{171}{640} c_1 \Lambda ^{2/3} n^2 R^{4/3}
e^{\frac{3 \sqrt[3]{\Lambda } n}{4 \sqrt[3]{R}}}+\frac{3}{10} c_1
\sqrt[3]{\Lambda } n R^{5/3} e^{\frac{3 \sqrt[3]{\Lambda } n}{4
\sqrt[3]{R}}}+\frac{1}{2} c_1 R^2 e^{\frac{3 \sqrt[3]{\Lambda }
n}{4 \sqrt[3]{R}}}-\frac{81 c_1 \Lambda ^{5/3} n^5 \sqrt[3]{R}
e^{\frac{3 \sqrt[3]{\Lambda } n}{4 \sqrt[3]{R}}}}{81920}\\ \notag
& -\frac{27 c_1 \Lambda ^{4/3} n^4 R^{2/3} e^{\frac{3
\sqrt[3]{\Lambda } n}{4 \sqrt[3]{R}}}}{20480}-\frac{9 c_1 \Lambda
n^3 R e^{\frac{3 \sqrt[3]{\Lambda } n}{4 \sqrt[3]{R}}}}{2560}
\frac{243 c_1 \Lambda ^2 n^6 \, \text{Ei}\left(\frac{3 n
\sqrt[3]{\Lambda }}{4 \sqrt[3]{R}}\right)}{327680}-\frac{27}{128}
c_1 \Lambda  n^3 R \, \text{Ei}\left(\frac{3 n \sqrt[3]{\Lambda
}}{4 \sqrt[3]{R}}\right) \, .
\end{align}
\end{itemize}
In the case $k>\alpha$, the models that one obtains by solving the
equation $x=-n\,\left(\frac{R}{\Lambda}\right)^{-m}$ result to
complex functions, so one is required to use the equation
$x=n\,\left(\frac{R}{\Lambda}\right)^{-m}$, in order to have real
functions. In this case, the criterion (\ref{constraintx2}) is
violated. Let us quote one example of this sort, for example
$m=5/2$ in which case, the solution to the equation
$x=n\,\left(\frac{R}{\Lambda}\right)^{-5/2}$ is,
\begin{equation}\label{sol512123446565}
F(R)=c_3 R+c_2+\frac{1}{2} c_1 R^2 e^{-\frac{\Lambda ^{5/2} n}{10
R^{5/2}}}-\frac{c_1 R^2 \left(\frac{\Lambda ^{5/2}
n}{R^{5/2}}\right)^{2/5} \Gamma \left(\frac{3}{5},\frac{n \Lambda
^{5/2}}{10 R^{5/2}}\right)}{10^{2/5}}+\frac{c_1 R^2
\left(\frac{\Lambda ^{5/2} n}{R^{5/2}}\right)^{4/5} \Gamma
\left(\frac{1}{5},\frac{n \Lambda ^{5/2}}{10 R^{5/2}}\right)}{2\
10^{4/5}} \, .
\end{equation}
Notice again in Eq. (\ref{sol512123446565}) the sign in the
exponentials, which makes the late-time description impossible.

A common feature of the models we discussed in this section is
that primordially these are described by an $R^2$ gravity, which
is known to provide a unique quasi-de Sitter evolution. $R^2$
gravity enjoys an elevated role among all $F(R)$ gravities, once
quasi-de Sitter solutions are considered. It is worth recalling
this feature, in order to strengthen our result here. This special
role of the $R^2$ gravity among all $F(R)$ gravities was
highlighted in Ref. \cite{Odintsov:2017tbc} using a dynamical
systems approach. Let us recall it in brief, so by introducing the
following dimensionless variables in vacuum $F(R)$ gravity,
\begin{equation}\label{variablesslowdown}
x_1=-\frac{\dot{F_R}(R)}{F_R(R)H},\,\,\,x_2=-\frac{F(R)}{6F(R)H^2},\,\,\,x_3=
\frac{R}{6H^2}\, ,
\end{equation}
the $F(R)$ gravity field equations can be expressed in terms of an
autonomous dynamical system in the following way,
\begin{align}\label{dynamicalsystemmain}
& \frac{\mathrm{d}x_1}{\mathrm{d}N}=-4-3x_1+2x_3-x_1x_3+x_1^2\, ,
\\ \notag &
\frac{\mathrm{d}x_2}{\mathrm{d}N}=8+m-4x_3+x_2x_1-2x_2x_3+4x_2 \, ,\\
\notag & \frac{\mathrm{d}x_3}{\mathrm{d}N}=-8-m+8x_3-2x_3^2 \, ,
\end{align}
where $m$ is defined to be,
\begin{equation}\label{parameterm}
m=-\frac{\ddot{H}}{H^3}\, .
\end{equation}
When the parameter $m$ is constant, the dynamical system
(\ref{dynamicalsystemmain}) is autonomous. In the case of a
quasi-de Sitter evolution with the scale factor being $a(t)=e^{H_0
t-H_i t^2}$, the parameter $m$ is identically equal to zero. The
total EoS of the system is equal to \cite{reviews1},
\begin{equation}\label{weffoneeqn}
w_{eff}=-1-\frac{2\dot{H}}{3H^2}\, ,
\end{equation}
and expressed in terms of $x_3$ is written,
\begin{equation}\label{eos1}
w_{eff}=-\frac{1}{3} (2 x_3-1)\, .
\end{equation}
We can easily find the fixed points of the dynamical system Eq.
(\ref{dynamicalsystemmain}) with $m=0$, which are,
\begin{equation}\label{fixedpointdesitter}
\phi_*^1=(-1,0,2),\,\,\,\phi_*^2=(0,-1,2)\, ,
\end{equation}
and the corresponding eigenvalues of the linearized matrix which
corresponds to the dynamical system for $\phi_*^1$ are $(-1, -1,
0)$, while for $\phi_*^2$ are $(1, 0, 0)$. Thus, the dynamical
system (\ref{dynamicalsystemmain}) has a stable non-hyperbolic
fixed point, the fixed point $\phi_*^1$ and one unstable fixed
point, namely $\phi_*^2$. These two fixed points are de Sitter
fixed points with $w_{eff}=-1$, however the second fixed point,
namely $\phi_*^2=(0,-1,2)$ indicates that $x_1\simeq 0$ and
$x_2\simeq -1$ which indicate that,
\begin{align}\label{caseidiffseqns1}
-\frac{\mathrm{d}^2F}{\mathrm{d}R^2}\frac{\dot{R}}{H\frac{\mathrm{d}F}{\mathrm{d}R}}\simeq
0,\,\,\,-\frac{F}{H^2\frac{\mathrm{d}F}{\mathrm{d}R}6}\simeq -1\,
.
\end{align}
For a slow-roll era, we have,
\begin{equation}\label{seconddiff}
F\simeq \frac{\mathrm{d}F}{\mathrm{d}R} \frac{R}{2}\, ,
\end{equation}
thus finally we have,
\begin{equation}\label{approximatersquare}
F(R)\simeq \alpha R^2\, ,
\end{equation}
where $\alpha$ is an arbitrary integration constant. Thus $R^2$
gravity is related to the unstable quasi-de Sitter fixed point of
the whole de Sitter solutions subspace of the $F(R)$ gravity phase
space. This clearly shows the elevated role of $R^2$ gravity among
all $F(R)$ gravities, once quasi-de Sitter solutions are
considered.

Now, the new feature of the models we considered, which are
consistent with the scalaron criterion on monotonicity
(\ref{derivativescalaronmass}) and de Sitter criterion
(\ref{criterion3}) is that these models can provide an $R^2$
inflationary era, which is known to provide a unique quasi-de
Sitter evolution and at the same time, one has a consistent
description of the dark energy era, a feature which we demonstrate
in detail in a later section. The inherent scale in these models
is the cosmological constant, which emerges in a unique way via
the equation $x=-n\,\left(\frac{R}{\Lambda}\right)^{-m}$, by
simply requiring that the cosmological constant is contained as a
scale in the $F(R)$ gravity function. Then by requiring that the
scalaron mass is a monotonically increasing function of the Ricci
scalar, or zero, one has models that provide an $R^2$ inflationary
era, and at the same time one has the same $F(R)$ gravity
controlling in a successful way the dark energy era. Thus a
unified description of inflation and dark energy is achieved with
the same $F(R)$ gravity. Although such descriptions are known in
the literature \cite{Odintsov:2019evb,Oikonomou:2020qah}, this is
the first time that such a unified description is derived by first
principles based on the scalaron monotonicity and the existence of
a stable de Sitter solution. The full analysis of the dark energy
era for some of the models we discussed in this subsection will be
presented in a later section. Finally, let us note that the
scalaron mass for all the models of this section behaves in the
desired way, that is, at small curvatures, the scalaron mass is
small, and at large curvatures, the scalaron mass is large.

\subsubsection{Models with $x=-n\,\left(\frac{R}{\Lambda}\right)^{m}$ and $m\ll 1$: $\alpha$-attractor-like Inflation}

Now let us consider another scenario which might lead to a viable
$F(R)$ gravity inflation, namely cases which lead to
$x=-n\,\left(\frac{R}{\Lambda}\right)^{m}$ with $m\ll 1$ and
$0<n<1$. This is a perplexed situation since the $F(R)$ gravity is
not easy to tackle analytically, but the central theme of this
case is that if $m\ll 1$, one has
$\left(\frac{R}{\Lambda}\right)^{m}\sim 1$ and therefore $x\sim
-n$ in this case. Thus, this scenario yields a tensor-to-scalar
ratio of the form,
\begin{equation}\label{alphaattractorneequationnew}
r\sim 3\alpha (1-n_s)^2\, ,
\end{equation}
with $\alpha=\frac{16}{(4+n)^2}$, which is basically a sort of
$\alpha$-attractor inflation. Definitely the tensor-to-scalar
ratio is smaller in this case, compared to the $R^2$ inflation
one. Models of this sort may result for $m\sim 1/100$ for example,
but it is too hard to quote these models here, due to the length
of the resulting $F(R)$ gravity. We will give a simple example,
for $m=1/5$ since the behavior is similar for lower values of $m$.
For $m=1/5$, solving $x=-n\,\left(\frac{R}{\Lambda}\right)^{1/5}$
yields,
\begin{align}\label{alphatattractorfrgravity}
& F(R)=c_3 R+c_2+e^{-\frac{5 n \sqrt[5]{R}}{4 \sqrt[5]{\Lambda
}}}\Big{(}\frac{24576 c_1 \Lambda  R }{5 n^5}+\frac{24576 c_1
\Lambda ^{4/5} R^{6/5}}{25 n^4}+\frac{768 c_1 \Lambda ^{3/5}
R^{7/5}}{5 n^3}+\frac{16 c_1 \Lambda ^{2/5}
R^{8/5}}{n^2}+\frac{2378170368 c_1 \Lambda ^{8/5} R^{2/5}}{15625
n^8}\\
\notag & +\frac{198180864 c_1 \Lambda ^{7/5} R^{3/5} }{3125
n^7}+\frac{12386304 c_1 \Lambda ^{6/5} R^{4/5}}{625 n^6}
+\frac{76101451776 c_1 \Lambda ^2 }{390625
n^{10}}+\frac{19025362944 c_1 \Lambda ^{9/5} \sqrt[5]{R}}{78125
n^9}\Big{)}\, ,
\end{align}
which during inflation is a sum of power-law $F(R)$ gravities. It
is conceivable that the case $m=1/100$ contains much more terms
that the above. Now there is a caveat with this case, having to do
with the viability of the model, which cannot be checked easily.
Since the models of the form
$x=-n\,\left(\frac{R}{\Lambda}\right)^{m}$ with $m\ll 1$ yield a
large number of power-law terms during inflation, it is impossible
to evaluate in a closed form the first slow-roll index
$\epsilon_1$. Thus one must perform some numerical analysis toward
evaluating the first slow-roll index. But our analysis offers many
advantages since the only thing required to validate that the
inflationary phenomenology of a specific model is viable is the
first slow-roll index. Thus one may solve the field equations for
appropriate initial conditions for early times, and give an
estimate for the first slow-roll index. This can yield estimates
for the spectral index, and thus the phenomenology of the model
can be checked in a straightforward way, regardless the lack of
analyticity. This numerical analysis based method is quite
important, so we will devote an entire section later on in this
article. The same numerical analysis can be performed for other
forms of the parameter $x$, for example $x=-n\,
e^{-\frac{R}{\Lambda}}$, which results to the following $F(R)$
gravity,
\begin{equation}\label{aboveFR}
F(R)=c_3 R+c_2+\int_1^R c_1 (-\Lambda ) \,
\text{Ei}\left(\frac{1}{4} e^{-\frac{X}{\Lambda }} n \Lambda
\right) \, dX\, .
\end{equation}
Intuitively, one understands that the above $F(R)$ gravity is
similar to simple Einstein-Hilbert gravity primordially, thus it
is hard to describe inflation with it. However, the $F(R)$ gravity
(\ref{aboveFR}) yields a small $x$ primordially, thus it obscures
the whole analysis, nevertheless a numerical analysis will reveal
that such an $F(R)$ gravity will produce a large first slow-roll
index and thus cannot describe inflation at all. We will return to
the need of numerical analysis for some complex models in a later
section in this article.

In this section we demonstrated that the $F(R)$ gravity
description of $\alpha$-attractors is possibly in the form of a
large number of power-law $F(R)$ gravity terms. It should be noted
that it is nearly impossible to obtain directly from the Einstein
frame the $F(R)$ gravity description of $\alpha$-attractors, since
given a scalar $\alpha$-attractor potential, one needs to solve
analytically the following equation \cite{Odintsov:2020thl},
\begin{equation}\label{solvequation}
RF_R=2\sqrt{\frac{3}{2}}\frac{\mathrm{d}}{\mathrm{d}\varphi}\left(\frac{V(\varphi)}{e^{-2\left(\sqrt{2/3}\right)\varphi}}\right)
\end{equation}
with $F_R=\frac{\mathrm{d}F(R)}{\mathrm{d}R}$, which is impossible
to solve for general $\alpha$.

\subsection{An Important Class of Exponential $R^2$ Deformations}

There is an important class of $F(R)$ gravity models which leads
to a unified description of inflation and the dark energy era.
These models have the following simplified form,
\begin{equation}\label{simlifiedexponential}
F(R)=R+\frac{R^2}{M^2}+\lambda
R\,e^{\epsilon\left(\frac{\Lambda}{R}\right)^{\beta}}+\lambda
\Lambda n \epsilon\, ,
\end{equation}
with $\epsilon$, $\lambda$, $\beta$ and $n$ being dimensionless
parameters. This particular class of models yield,
\begin{equation}\label{xexp}
x\sim -\mathcal{C}\,\frac{M^2\Lambda^{\beta}}{R\,R^{\beta}}
\end{equation}
in the large curvature regime during the inflationary era, with
$\mathcal{C}=2\beta  \left(\beta ^2-1\right) \lambda  \epsilon$,
thus $x\sim 0$ and the $R^2$ term dominates the evolution during
the inflationary era. More importantly, these models also yield a
viable dark energy era as we will demonstrate in a later section,
and specifically we will show that $\Omega_{DE}(0)=0.6901$
regarding the dark energy density parameter, while the dark energy
EoS parameter is $\omega_{DE}(0)=-1.036$ for $\beta=0.99$
$\lambda=0.8$, $\epsilon=9.1$ and $n=0.099$. The exceptional class
of exponential deformations of the $R^2$ model stem naturally from
the requirements that the de Sitter mass is a monotonic function
of the Ricci scalar and also that $x$ is almost zero.

\subsection{Alternative Viable Models}

Let us quote several other models which can describe inflation and
dark energy in a unified way, and also the models are compatible
with the de Sitter scalaron mass positivity (\ref{criterion1}) and
the monotonicity criterion (\ref{derivativescalaronmass}). All
these models yield quite interesting phenomenology since the de
Sitter scalaron mass is always positive, both at the early and
late-time de Sitter eras. These models were developed in Ref.
\cite{Oikonomou:2022wuk}, and yield different in the resulting
functional form of the parameter $x$, however their phenomenology
is quite similar. Let us start with the model,
\begin{equation}\label{fr22}
    F(R)=R+\frac{R^2}{M^2}-\frac{\beta \Lambda}{c+\log(\epsilon\,R/m_s^2)},
\end{equation}
Primordially, this models is described by an $R^2$ gravity, but at
late-times the last term dominates and a viable dark energy era is
accomplished by choosing $\beta=0.5\,, c=1 , \epsilon=1/220$.
Specifically, regarding the late-time phenomenology, one gets,
$\Omega_{DE}(0)=0.6834$ regarding the dark energy density
parameter, while the dark energy EoS parameter is
$\omega_{DE}(0)=-1.0372$, which are compatible with the Planck
data on the cosmological parameters $\Omega_{DE}=0.6847 \pm
0.0073$ and $\omega_{DE}=-1.018 \pm 0.031$. This models stems from
a $x$ parameter of the form,
\begin{equation}\label{xmodel1extra}
x=-\frac{8 \beta  \Lambda  M^2 \left(\log \left(\frac{R \epsilon
}{m_s^2}\right) \left(\log \left(\frac{R \epsilon
}{m_s^2}\right)+5\right)+7\right)}{\left(\log \left(\frac{R
\epsilon }{m_s^2}\right)+1\right) \left(3 \beta  \Lambda M^2+\log
\left(\frac{R \epsilon }{m_s^2}\right) \left(\beta \Lambda M^2+2
R^2 \log \left(\frac{R \epsilon }{m_s^2}\right) \left(\log
\left(\frac{R \epsilon }{m_s^2}\right)+3\right)+6 R^2\right)+2
R^2\right)}\, .
\end{equation}
Now it can easily be checked  that the parameter $x$ is negative
and very small, in fact, $x\sim 0$ in the large curvature regime.
This can only be done numerically, by choosing sensible values for
the curvature during inflation, like in Eq. (\ref{roverlambda}).
Another viable model with perplexed form of the parameter $x$ is
the following,
\begin{equation}\label{fr221}
    F(R)=R+\frac{R^2}{M^2}-\frac{\beta  \Lambda }{\gamma +\frac{1}{\log \left(\frac{R \epsilon }{m_s^2}\right)}},
\end{equation}
As in the previous model, this model is primordially described by
an $R^2$ gravity, but at late-times the last term dominates and a
viable dark energy era is accomplished by choosing $\beta=11.81\,,
\gamma=1.5 , \epsilon=100$. Specifically, regarding the late-time
phenomenology for this model, one gets, $\Omega_{DE}(0)=0.6876$
regarding the dark energy density parameter, while the dark energy
EoS parameter is $\omega_{DE}=-0.9891$,  which are compatible with
the Planck data on the cosmological parameters. This models stems
from a $x$ parameter of the form,
\begin{equation}\label{xmodel1extra1}
x=-\frac{8 \beta  \Lambda  M^2 \left(3 \gamma ^2+3 \gamma +\gamma
^2 \log ^2\left(\frac{R \epsilon }{m_s^2}\right)+(3 \gamma +2)
\gamma  \log \left(\frac{R \epsilon
}{m_s^2}\right)+1\right)}{\left(\gamma  \log \left(\frac{R
\epsilon }{m_s^2}\right)+1\right) \left(\beta  (2 \gamma +1)
\Lambda  M^2+\gamma  \left(\beta  \Lambda  M^2+6 R^2\right) \log
\left(\frac{R \epsilon }{m_s^2}\right)+2 \gamma ^3 R^2 \log
^3\left(\frac{R \epsilon }{m_s^2}\right)+6 \gamma ^2 R^2 \log
^2\left(\frac{R \epsilon }{m_s^2}\right)+2 R^2\right)}\, .
\end{equation}
Now it can easily be checked  that in this case too, the parameter
$x$ negative and very small, in fact, $x\sim 0$ in the large
curvature regime. Let us quote here another viable model with
perplexed form of the parameter $x$, the following,
\begin{equation}\label{fr2212}
    F(R)=R+\frac{R^2}{M^2}-\frac{\beta  \Lambda }{\gamma +\exp \left(-\frac{R \epsilon }{m_s^2}\right)},
\end{equation}
As in the previous models, this model is also primordially
described by an $R^2$ gravity, but at late-times the last term
dominates and a viable dark energy era is accomplished by choosing
$\beta=20\,, \gamma=2 , \epsilon=0.00091$. Specifically, regarding
the late-time phenomenology for this model, we have,
$\Omega_{DE}(0)=0.6918$ regarding the dark energy density
parameter, while the dark energy EoS parameter is
$\omega_{DE}=-0.9974$,  which are compatible with the Planck data
on the cosmological parameters. This models stems from a $x$
parameter of the form,
\begin{equation}\label{xmodel1extra12}
x=-\frac{4 \beta  \Lambda  M^2 R \epsilon ^3 e^{\frac{R \epsilon
}{m_s^2}} \left(\gamma  e^{\frac{R \epsilon }{m_s^2}} \left(\gamma
e^{\frac{R \epsilon }{m_s^2}}-4\right)+1\right)}{m_s^2 \left(\beta
\Lambda M^2 \epsilon ^2 e^{\frac{R \epsilon }{m_s^2}} \left(\gamma
^2 e^{\frac{2 R \epsilon }{m_s^2}}-1\right)+2 \left(\gamma m_s
e^{\frac{R \epsilon }{m_s^2}}+m_s\right)^4\right)}\, .
\end{equation}
Now it can easily be checked  that in this case too, the parameter
$x$ negative and very small, in fact, $x\sim 0$ in the large
curvature regime. Another viable model with a peculiar form of the
parameter $x$, has the following form,
\begin{equation}\label{fr22124}
    F(R)=R+\frac{R^2}{M^2}-\frac{\beta  \Lambda  \left(\frac{R}{m_s^2}\right)^n}{\delta +\gamma  \left(\frac{R}{m_s^2}\right)^n},
\end{equation}
As in the previous models, this model is also primordially an
$R^2$ gravity, but at late-times the last term dominates again,
and a viable dark energy era is accomplished by choosing
$\beta=1.4\,, \gamma=0.2,\, \epsilon=0.00091\,
,\delta=0.2,\,n=0.3$. Specifically, regarding the late-time
phenomenology for this model, we get, $\Omega_{DE}(0)=0.6851$
regarding the dark energy density parameter, while the dark energy
EoS parameter is $\omega_{DE}=-0.9887$,  which are again
compatible with the Planck data on the cosmological parameters.
This models stems from a $x$ parameter of the form,
\begin{equation}\label{xmodel1extra124}
x=\frac{4 \beta  \delta  \Lambda  M^2 n
\left(\frac{R}{m_s^2}\right)^n \left(4 \gamma  \delta
\left(n^2-1\right) \left(\frac{R}{m_s^2}\right)^n-\gamma ^2 (n+1)
(n+2) \left(\frac{R}{m_s^2}\right)^{2 n}-\delta ^2 (n-2)
(n-1)\right)}{\beta  \delta  \Lambda  M^2 n
\left(\frac{R}{m_s^2}\right)^n \left(\delta +\gamma
\left(\frac{R}{m_s^2}\right)^n\right) \left(\delta +\gamma (n+1)
\left(\frac{R}{m_s^2}\right)^n-\delta  n\right)+2 R^2 \left(\delta
+\gamma \left(\frac{R}{m_s^2}\right)^n\right)^4}\, .
\end{equation}
Now it can easily be checked  that in this case too, the parameter
$x$ negative and very small, in fact, $x\sim 0$ in the large
curvature regime. Finally, let us quote a last model with
perplexed form of the parameter $x$, the following,
\begin{equation}\label{fr22125}
    F(R)=R+\frac{R^2}{M^2}-\Lambda  \left(\gamma -\exp \left(-\frac{R \epsilon }{m_s^2}\right)\right),
\end{equation}
As in the previous models, this model is also primordially
described by an $R^2$ gravity, but at late-times the last term
dominates and a viable dark energy era is accomplished by choosing
$\gamma=7.5 , \epsilon=0.0005$. Specifically, regarding the
late-time phenomenology in this case, one obtains,
$\Omega_{DE}(0)=0.6847$ regarding the dark energy density
parameter, and the dark energy EoS parameter is
$\omega_{DE}=-1.0367$, which are again compatible with the Planck
data on the cosmological parameters. This models stems from a
quite simple $x$ parameter of the form,
\begin{equation}\label{xmodel1extra125}
x=-\frac{4 \Lambda  M^2 R \epsilon ^3}{\Lambda  M^2 m_s^2 \epsilon
^2+2 m_s^6 e^{\frac{R \epsilon }{m_s^2}}}\, .
\end{equation}
Now it can easily be checked that in the large curvature regime
one has for this model,
\begin{equation}\label{xapprxolast}
x\sim -\frac{2 \Lambda  M^2 R \epsilon ^3 e^{-\frac{R \epsilon
}{m_s^2}}}{m_s^6}\, ,
\end{equation}
which is negative and almost zero. All the models we describe here
have some interesting characteristics that all the viable models
of this section share:
\begin{itemize}
    \item All the models result to a unification of early and late-time
acceleration.
    \item All the models yield primordially $x$ in the range $-1\leq x\leq
    0$, and in fact $x\sim 0$ and negative.
    \item All the models have positive de Sitter scalaron mass
    both at early and late times and also the de Sitter scalaron
    mass is primordially small, while at late times it is large.
\end{itemize}
Now it is not certain that every model which yields a parameter
$-1\leq x\leq 0$ will be viable, but all the viable models which
unify early and late-time acceleration, do yield $-1\leq x\leq 0$.
This is a clear indication of a pattern for viable models that
provide a unified description of inflation and the dark energy
era. Also models which can probably yield a viable inflation might
lead to a parameter $x>0$. One example of this sort is a slight
deformation of the $R^2$ model,
\begin{equation}\label{antiexample}
F(R)=R+\frac{R^{\epsilon +2}}{M^2}\, ,
\end{equation}
with $\epsilon\ll 1$. This model yields $x=4\epsilon$, which when
$\epsilon>0$, it is positive. But in this case $x$ is constant, so
this case cannot be dealt with the formalism developed in the
previous section and used in this section. The case $x=$const will
be dealt in a later section, and clearly cannot describe inflation
and dark energy in a unified way. Another $R^2$ which may yield a
viable inflationary era is,
\begin{equation}\label{erfsdfgr}
F(R)=R+\frac{R^2}{M^2} \log \left(\frac{R}{\Lambda }\right)\, ,
\end{equation}
which yields a positive $x=\frac{8}{2 \log \left(\frac{R}{\Lambda
}\right)+3}$, however this case might yield a negative de Sitter
scalaron mass, since one has $m^2(R)=\frac{M^2-2 R}{6 \log
\left(\frac{R}{\Lambda }\right)+9}$. In addition, this
inflationary model cannot describe inflation and dark energy in a
unified way.

\section{Non-viable Scenarios: Non-de Sitter Solutions in $F(R)$ Gravity}

A this point we shall consider scenarios which yield a large $x$
parameter at first horizon crossing, and thus are essentially
non-viable since the spectral index of the scalar perturbations
becomes too large to be compatible with a nearly scale invariant
power spectrum. These models are consist of any model that can
yield a large $x$ parameter, including models of the form:
\begin{itemize}
    \item $x\sim \ln (\frac{R}{\Lambda})$,
    \item $x\sim e^{\frac{R}{\Lambda}}$,
    \item $x\sim \left(\frac{R}{\Lambda}\right)^{m}\,\ln (\frac{R}{\Lambda})$,
    $m>0$,
    \item $x\sim \left(\frac{R}{\Lambda}\right)^{m}\,e^{\frac{R}{\Lambda}}$,
    $m>0$,
    \item $x\sim \left(\frac{R}{\Lambda}\right)^{m}$, $m>0$,
\end{itemize}
or any other combination of functions that can yield a large
(infinite) parameter $x$ at first horizon crossing. From the
above, only the form $x=n\,\left(\frac{R}{\Lambda}\right)^{m}$
yields analytical results, with $n$ some arbitrary number, the
sign of which plays no essential role. It must be mentioned that
models which yield $x\ll-1$ or even $x>0$ are also non-viable
since these violate the scalaron mass monotonicity criterion
(\ref{constraintx2}). We will concentrate on the models
$x=n\,\left(\frac{R}{\Lambda}\right)^{m}$, with $m$ some positive
integer or rational number. Let us start with the integer cases
first so we will examine some characteristic cases, with
$m=1,2,3,4,5$ which yield some results in closed form. Let us
start with the case $m=1$ first, in which case, by solving
$x=-n\,\left(\frac{R}{\Lambda}\right)$, we obtain analytically the
following solution,
\begin{equation}\label{sol1nonviable}
F(R)=c_3 R+c_2+\frac{16 c_1 \Lambda ^2 e^{-\frac{n R}{4 \Lambda
}}}{n^2}\, ,
\end{equation}
where $c_i$, $i=1,2,3$ are integration constants, while the
equation $x=n\,\left(\frac{R}{\Lambda}\right)$ yields,
\begin{equation}\label{sol1nonviable1}
F(R)=c_3 R+c_2+\frac{16 c_1 \Lambda ^2 e^{\frac{n R}{4 \Lambda
}}}{n^2}\, ,
\end{equation}
with the first case (\ref{sol1nonviable}) being some
Einstein-Hilbert gravity during inflation, while the second case
(\ref{sol1nonviable1}) being essentially an exponential model.
With our method, clearly these models are non-viable which is a
valuable result since the inflationary phenomenology of these
models cannot be dealt analytically.

Now let us proceed to the case $m=2$, in which case, by solving
$x=-n\,\left(\frac{R}{\Lambda}\right)^2$, we obtain analytically
the following solution,
\begin{equation}\label{sol1nonviablem2}
F(R)=c_3 R+c_2+\frac{\sqrt{2 \pi } c_1 \Lambda  \left(R \,
\text{erf}\left(\frac{\sqrt{n} R}{2 \sqrt{2} \Lambda
}\right)+\frac{2 \sqrt{\frac{2}{\pi }} \Lambda  e^{-\frac{n R^2}{8
\Lambda ^2}}}{\sqrt{n}}\right)}{\sqrt{n}}\, ,
\end{equation}
where $c_i$, $i=1,2,3$ are integration constants, while the
equation $x=n\,\left(\frac{R}{\Lambda}\right)^2$ yields,
\begin{equation}\label{sol1nonviable1m2}
F(R)=c_3 R+c_2+\frac{\sqrt{2 \pi } c_1 \Lambda  \left(R \,
\text{erfi}\left(\frac{\sqrt{n} R}{2 \sqrt{2} \Lambda
}\right)-\frac{2 \sqrt{\frac{2}{\pi }} \Lambda  e^{\frac{n R^2}{8
\Lambda ^2}}}{\sqrt{n}}\right)}{\sqrt{n}}\, ,
\end{equation}
with the first case (\ref{sol1nonviablem2}) being again some
Einstein-Hilbert gravity during inflation, while the second case
(\ref{sol1nonviable1m2}) being again an exponential model. Our
method is proven valuable since both the models quoted above are
deemed non-viable without getting into detailed calculations.

Now let us proceed to the case $m=3$, in which case, only the case
$x=-n\,\left(\frac{R}{\Lambda}\right)^3$ can yield real $F(R)$
gravity forms. So by solving
$x=-n\,\left(\frac{R}{\Lambda}\right)^3$, we obtain analytically
the following solution,
\begin{equation}\label{sol1nonviablem2m3}
F(R)=c_3 R+c_2-\left(\frac{2}{3}\right)^{2/3} c_1 \left(\frac{R^2
\Gamma \left(\frac{1}{3},\frac{n R^3}{12 \Lambda
^3}\right)}{\sqrt[3]{\frac{n R^3}{\Lambda ^3}}}-\frac{2^{2/3}
\sqrt[3]{3} R^2 \Gamma \left(\frac{2}{3},\frac{n R^3}{12 \Lambda
^3}\right)}{\left(\frac{n R^3}{\Lambda ^3}\right)^{2/3}}\right)\,
,
\end{equation}
while for $m=4$ we obtain,
\begin{equation}\label{sol1nonviablem2m3}
F(R)=c_3 R+c_2-\frac{1}{2} c_1 \left(\frac{2 \sqrt{\pi } \Lambda
^2 \, \text{erf}\left(\frac{\sqrt{n} R^2}{4 \Lambda
^2}\right)}{\sqrt{n}}+\frac{R^2 \Gamma \left(\frac{1}{4},\frac{n
R^4}{16 \Lambda ^4}\right)}{\sqrt[4]{\frac{n R^4}{\Lambda
^4}}}\right)\, ,
\end{equation}
and for $m=5$ we get,
\begin{equation}\label{sol1nonviablem2m3m5}
F(R)=c_3 R+c_2-\frac{2^{2/5} c_1 \left(\frac{R^2 \Gamma
\left(\frac{1}{5},\frac{n R^5}{20 \Lambda
^5}\right)}{\sqrt[5]{\frac{n R^5}{\Lambda ^5}}}-\frac{2^{2/5}
\sqrt[5]{5} R^2 \Gamma \left(\frac{2}{5},\frac{n R^5}{20 \Lambda
^5}\right)}{\left(\frac{n R^5}{\Lambda
^5}\right)^{2/5}}\right)}{5^{4/5}}\, .
\end{equation}
We need to note that for $m>5$ and $m$ integer the behavior of the
solutions to the equation $x=-n\,\left(\frac{R}{\Lambda}\right)^m$
is functionally similar to the solution
(\ref{sol1nonviablem2m3m5}), with the powers of the curvature
changing of course. Now let us consider the cases for which $m$ is
some rational number $m=\frac{k}{\alpha}$, with $k$ and $\alpha$
some positive integers. Let us consider the cases $k<\alpha$, and
let us focus on the case $m=1/2$ firstly, so by solving the
equation $x=-n\,\left(\frac{R}{\Lambda}\right)^{1/2}$ we get,
\begin{equation}\label{sol1nonviablem2m12a}
F(R)=c_3 R+c_2-\frac{8 c_1 \Lambda ^2 e^{-\frac{1}{2} n
\sqrt{\frac{R}{\Lambda }}} \left(-\frac{24}{n^2}-\frac{12
\sqrt{\frac{R}{\Lambda }}}{n}-\frac{2 R}{\Lambda }\right)}{n^2}\,
,
\end{equation}
while the equation $x=n\,\left(\frac{R}{\Lambda}\right)^{1/2}$
yields,
\begin{equation}\label{sol1nonviable1m212}
F(R)=c_3 R+c_2+\frac{8 c_1 \Lambda ^2 e^{\frac{1}{2} n
\sqrt{\frac{R}{\Lambda }}} \left(\frac{24}{n^2}-\frac{12
\sqrt{\frac{R}{\Lambda }}}{n}+\frac{2 R}{\Lambda }\right)}{n^2}\,
,
\end{equation}
with both cases (\ref{sol1nonviablem2m12a}) and
(\ref{sol1nonviable1m212}) being some $R^{1/2}$ exponential
containing gravities. Our method is proven valuable since both the
models quoted above are again deemed non-viable without getting
into detailed calculations. Now let us consider other cases with
$m=\frac{k}{\alpha}$, and with $k<\alpha$, so let us consider
$m=5/6$ and solve $x=-n\,\left(\frac{R}{\Lambda}\right)^{5/6}$, we
get,
\begin{align}\label{sol1nonviable1m212234}
& F(R)=c_3 R+c_2+\frac{56 c_1 \Lambda ^{5/3} \sqrt[3]{R}
e^{-\frac{3 n R^{5/6}}{10 \Lambda ^{5/6}}}}{3 n^2}-\frac{4
\sqrt[5]{\frac{2}{3}} c_1 \Lambda ^{5/3} \sqrt[3]{R} \left(\frac{n
R^{5/6}}{\Lambda ^{5/6}}\right)^{4/5} \Gamma
\left(\frac{1}{5},\frac{3 n R^{5/6}}{10 \Lambda
^{5/6}}\right)}{5^{4/5} n^2}\\ \notag &+\frac{112
\left(\frac{2}{3}\right)^{2/5} c_1 \Lambda ^{5/2} \left(\frac{n
R^{5/6}}{\Lambda ^{5/6}}\right)^{3/5} \Gamma
\left(\frac{2}{5},\frac{3 n R^{5/6}}{10 \Lambda ^{5/6}}\right)}{3\
5^{3/5} n^3 \sqrt{R}} \, ,
\end{align}
while the solution for $x=n\,\left(\frac{R}{\Lambda}\right)^{5/6}$
yields complex functional forms for the $F(R)$ gravity.
Considering a scenario with $m=\frac{k}{\alpha}$, and with
$k>\alpha$ and specifically, $m=7/3$, by solving
$x=-n\,\left(\frac{R}{\Lambda}\right)^{7/3}$, we get,
\begin{align}\label{sol1nonviable1m212234cheat}
& F(R)=c_3 R+c_2-\frac{2 \sqrt[7]{\frac{3}{7}} 2^{5/7} c_1 \Lambda
^{7/3} \sqrt[7]{\frac{n R^{7/3}}{\Lambda ^{7/3}}} \Gamma
\left(\frac{6}{7},\frac{3 n R^{7/3}}{28 \Lambda ^{7/3}}\right)}{n
\sqrt[3]{R}}-\frac{\left(\frac{3}{7}\right)^{4/7} 2^{6/7} c_1
\Lambda ^{7/3} \left(\frac{n R^{7/3}}{\Lambda ^{7/3}}\right)^{4/7}
\Gamma \left(\frac{3}{7},\frac{3 n R^{7/3}}{28 \Lambda
^{7/3}}\right)}{n \sqrt[3]{R}} \, .
\end{align}
These are quite complex forms of $F(R)$ gravity, which without our
proposed method could not be deemed viable or non-viable easily,
unless approximations were used. Now with our method, these are
easily deemed non viable since these lead to an unacceptable value
for the parameter $x$ at first horizon crossing, thus there is no
reason in evaluating the first slow-roll index.

We presented some non-viable cases of $F(R)$ gravity using a very
simple approach, which reduces simply in evaluating the parameter
$x$ defined in Eq. (\ref{parameterxfinal}), namely, $x=\frac{4
F_{RRR}\,R}{F_{RR}}$. The cases of non-viability at first horizon
crossing are dictated by the following cases:
\begin{itemize}
    \item If the scalaron mass monotonicity criterion (\ref{derivativescalaronmass}) is
    violated,which occurs when $x<-1$ or $x>0$, that is $\lim_{\frac{R}{\Lambda}\to \infty}x>0$ or $\lim_{\frac{R}{\Lambda}\to \infty}x<-1$ at
    first horizon
    crossing, then the $F(R)$ gravity model is deemed non-viable.

    \item If $\lim_{\frac{R}{\Lambda}\to \infty}x\to \infty$, then
    the $F(R)$ gravity is non viable.
\end{itemize}
We believe it is the first time that such a simple and
straightforward technique has been given for $F(R)$ gravity
inflation, and in this section we presented several models that
emerged by solving analytically the differential equation
$x=\beta(R,\Lambda)$, and we showed that these models result to
non-viable $F(R)$ gravity inflation.

\section{Constant $x$ Scenarios: Disentangling Power-law Inflation in $F(R)$ Gravity from Power-law Evolution}

In this section we shall consider the scenario in which the
parameter $x$ is constant, which respect the de Sitter criterion
(\ref{constraintx2}). In the case that $x=-n$ (but in principle
one can have $x=n$, which we discuss at the end of this section),
where $0<n<1$, the solution of the equation $x=-n$ is equal to,
\begin{equation}\label{constnatxol}
F(R)=c_1+c_3 R+c_2\frac{16 c_1 R^{2-\frac{n}{4}}}{(n-8) (n-4)}\, ,
\end{equation}
which is clearly a power-law gravity. In the standard $F(R)$
gravity literature, power-law gravity $\sim R^p$ is known to yield
a power-law evolution of the for $H\sim 1/(p\,t)$. However in this
section we shall reveal a caveat in the standard approaches in
power-law gravities, and we shall analyze power-law gravity
inflation using our formalism developed in this article. Firstly,
let us demonstrate the caveat in the standard power-law $F(R)$
gravity that is used in the literature, we use a slightly
different notation since we deviate from the approach we adopted
in this work. Let us consider,
\begin{equation}\label{polynomialfr} f(R)=R+\beta R^n\, ,
\end{equation}
for any real $n$. The Friedman equation of the vacuum $f(R)$
gravity is,
\begin{equation}\label{friedmannewappendix}
3 H^2F=\frac{RF-f}{2}-3H\dot{F}\, ,
\end{equation}
with $F=\frac{\partial f}{\partial R}$. During the inflationary,
by $F\sim n\beta R^{n-1}$ hence the Friedman equation
(\ref{friedmannewappendix}) takes the form,
\begin{align}\label{eqnsofmkotionfrpolyappendix}
& 3 H^2n\beta R^{n-1}=\frac{\beta (n-1)R^{n-1}}{2}-3n(n-1)\beta
HR^{n-2}\dot{R}\, .
\end{align}
Now the standard approach in $F(R)$ gravity literature, takes into
account the slow-roll approximation, according to which, the Ricci
scalar $R=12H^2+6\dot{H}$ during inflation becomes at leading
order $R\sim 12 H^2$ and $\dot{R}\sim 24H\dot{H}$, therefore
Friedman equation (\ref{eqnsofmkotionfrpolyappendix}) takes the
following form,
\begin{equation}\label{leadingordereqnappendix}
3H^2n\beta \simeq 6\beta (n-1)H^2-6n\beta(n-1)\dot{H}+3\beta
(n-1)\dot{H}\, ,
\end{equation}
which can be solved analytically to yield,
\begin{equation}\label{hubblefrpolyappendix}
H(t)=\frac{1}{p\,t}\, ,
\end{equation}
where $p=\frac{n-2}{-2n^2+3n-1}$, which describes a power-law
evolution. This power-law evolution describes an inflationary era
when $1.36<n<2$. Let us now point out the problems of this
approach. Firstly, the power-law evolution
(\ref{hubblefrpolyappendix}) was derived under the assumption
$\dot{H}\ll H^2$, however for the power-law evolution one has
$\dot{H}=-p\,H^2$. For $n=1.37$ one has $p=-0.978565$, so clearly
the slow-roll condition is violated. Notably, the power-law
solution (\ref{hubblefrpolyappendix}) was derived using the
slow-roll assumption, so there is a big conflict in this approach.
The second caveat is that the solution $n=2$ does not describe
inflation in this context. Although nothing restricts $n$, apart
from the inflation evolution requirement, the $n=2$ solution
should describe inflation too, it is the $R^2$ model. Thus the
standard approach for power-law $F(R)$ gravity is problematic.

Note that in the standard power-law $F(R)$ inflation, the
slow-roll indices are,
\begin{equation}\label{epsiloniforfrpoly}
\epsilon_1=\frac{n-2}{1-3n+2n^2},\,\,\,\epsilon_3=(n-1)\epsilon_1,\,\,\,\epsilon_4=\frac{n-2}{n-1}\,
,
\end{equation}
and the observational indices are,
\begin{equation}\label{vacuumspectral}
n_s=1-6\epsilon_1-2\epsilon_4,\,\,\,r=48 \epsilon_1^2\, .
\end{equation}
One value of $n$ which yields a viable $n_s$ is $n=1.81$, however
the corresponding tensor-to-scalar ratio takes the value $r=0.13$
thus the power-law $F(R)$ gravity model is not compatible with the
Planck data \cite{Planck:2018jri}.

\begin{figure}
\centering
\includegraphics[width=18pc]{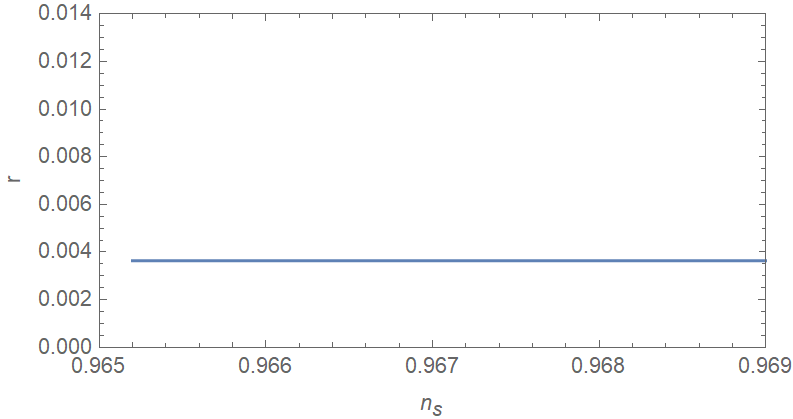}
\caption{The parametric plot of $n_s$ and $r$ for
$\epsilon_1=0.0087$ and $n=[0,0.5]$.}\label{plotparametricnsr}
\end{figure}

As we indicated, the standard approach in power-law $F(R)$ gravity
results to theoretical inconsistencies. Thus at this point let us
disentangle the power-law evolution from power-law gravity. Note
that the power-law evolution results to theories for which
$\dot{\epsilon}_1=0$, so from this point, let us assume that
power-law gravities do not lead to power-law evolutions of the
form $H\sim 1/(p\,t)$, which is clearly justified from the above.
Thus the formalism of the previous sections apply, therefore
power-law $F(R)$ gravities of the form (\ref{constnatxol}) are
obtained for $x=-n$, and this includes the $R^2$ gravity. Thus for
$0<n<1$, the resulting inflationary theory can in principle be
compatible with the Planck data. However in order to be formal,
one needs to calculate the first slow-roll index, and this is not
an easy task. Notice however that the solution (\ref{constnatxol})
is basically a deformation of the $R^2$ model. In order to have an
idea on which values $n$ can take in order to get viability with
the Planck data, we assume that the first slow-roll index takes
the value it has for the $R^2$ model, so $\epsilon_1\sim 0.0087$.
One easily obtains that for $0<n<0.5$, one obtains a viable
evolution. This can also be seen in Fig. \ref{plotparametricnsr}
where we present the parametric plot of $n_s$ and $r$ for
$\epsilon_1=0.0087$ and $n=[0,0.5]$.
Of course our approach is rather heuristic, and should in
certainly be valid for small values of $n$. Nevertheless, this is
actually a criterion for the validation of our approach. For $n$
chosen in the range $n=[0,0.5]$, the power-law gravity solution
interpolates between an $R^{1.875}$ and an $R^2$ gravity. So the
viable power-law $F(R)$ gravities are of the form
(\ref{constnatxol}) for $n=[0,0.5]$. This is an important outcome
of this work, where we disentangled power-law evolution from
power-law $F(R)$ gravity. As we showed, power-law $F(R)$ gravity
can in fact be viable and related to possibly a quasi-de Sitter
evolution. Another argument that provides solid proof for the
validity of our approach is when $n\sim 0$. Then, the power-law
gravity (\ref{constnatxol}) is a slight deformation of $R^2$
gravity which is known to be viable and is related to a quasi-de
Sitter evolution. Using the formalism of power-law $F(R)$ gravity
leading to a power-law solution (\ref{hubblefrpolyappendix})
 would render these $R^2$ deformations non-viable, a result which
 is clearly wrong. Nevertheless, in order to be correct formally,
 when $n\to 0.5$ one needs to implement a numerical calculation in
 order to determine the order of the first slow-roll index. This
 numerical method approach will be discussed in a later section.
 But our point is clear, small $R^2$ deformations are non-viable
 in standard $F(R)$ gravity formalism appearing in the literature,
 a result which is clearly wrong, however in our theoretical
 framework, these power-law $R^2$ deformations find an elegant and
 valid description. Also let us note that a power-law evolution
 would originate from theories which have $\dot{\epsilon}_1=0$
 during inflation. These theories cannot be generated using our
 formalism, and will be studied in a future work focused on this
 issue. Our point so far is clear, the viability of any $F(R)$
 gravity model can be determined by evaluating the parameter $x$
 and the value of the first slow-roll index at first horizon
 crossing. The latter might be difficult to be evaluated for a
 complex $F(R)$ gravity. However, during the slow-roll era when $R\sim 12
 H^2$ and also $\dot{H}\ll H^2$, one may have an approximate
 relation for the first slow-roll index for any $F(R)$ gravity,
 using the Friedmann equation. As we show in a later section, the first slow-roll index during the
 slow-roll era can be approximately equal to,
\begin{equation}\label{approxepsilon1}
\epsilon_1\sim \frac{2 F(R)-F_R R}{2 F_{RR}R^2}\, .
\end{equation}
So for the case at hand, for the $F(R)$ gravity of Eq.
(\ref{constnatxol}), we approximately have,
\begin{align}\label{approximateepsilon1powerlaw}
& \epsilon_1\sim \frac{2 n}{(n-8) (n-4)}+\frac{32 c_2
R^{\frac{n}{4}-2}}{c_1 (n-8) (n-4)}-\frac{12 c_2 n
R^{\frac{n}{4}-2}}{c_1 (n-8) (n-4)}\\ \notag & \frac{c_3 n^2
R^{\frac{n}{4}-1}}{2 c_1 (n-8) (n-4)}++\frac{c_2 n^2
R^{\frac{n}{4}-2}}{c_1 (n-8) (n-4)}+\frac{16 c_3
R^{\frac{n}{4}-1}}{c_1 (n-8) (n-4)}-\frac{6 c_3 n
R^{\frac{n}{4}-1}}{c_1 (n-8) (n-4)}\, ,
\end{align}
so at leading order, one has,
\begin{equation}\label{epsilonleading}
\epsilon_1\sim \frac{2 n}{(n-8) (n-4)}\, .
\end{equation}
Hence, as it can be checked, for $0.1<n<0.13$ one has a small
first slow-roll index, with a value similar to the one of the
$R^2$ model. For $n\ll 1$, one has an exact $R^2$ model at leading
order, so this is just a slight $R^2$ deformation inflation. Of
course this is not an exactly accurate technique, but it gives us
a hint on the values of the first slow-roll index. In a later
section we shall provide some details on this approximate
technique. The results for the power-law $F(R)$ gravity are
supported by the findings of Ref. \cite{Martin:2013tda}, which
also prove that power-law deformations of $R^2$ inflation, like
the viable cases we discussed, are indeed viable. This result
cannot be obtained by following the standard Jordan frame
treatment of $F(R)$ gravity which leads to power-law type
evolution.

Before closing, let us briefly consider another case of interest,
namely for $x=4$. This is clearly violating the de Sitter
constraint (\ref{constraintx2}), so it does not describe a viable
inflationary era. However, for $x=4$ Eq. (\ref{asxeto1}) yields
$n_s=1$, which describes a scale invariant evolution. This scale
invariant evolution is basically generated by an $R^3$ gravity,
which is easily found by solving $x=4$, and we have,
\begin{equation}\label{rcubescaleinv}
 F(R)=c_3 R+c_2+\frac{c_1 R^3}{6}\, .
 \end{equation}
Also let us note that the constant $x$ models clearly cannot
describe the dark energy era and the inflationary era in a unified
way, possibly only inflation. Furthermore, the case $x=n$ with
$n\ll 1$ and positive, also describes a viable inflationary era
since it is a slight deformation of $R^2$ gravity. Hence, the
criterion for the monotonicity of the de Sitter scalaron mass is
not expected to make sense in the case that $x$=const.

\section{A Unified Description of the Inflationary and Dark Energy Eras with $F(R)$ Gravity: Connecting Inflation and Dark Energy}

\begin{figure}
\centering
\includegraphics[width=18pc]{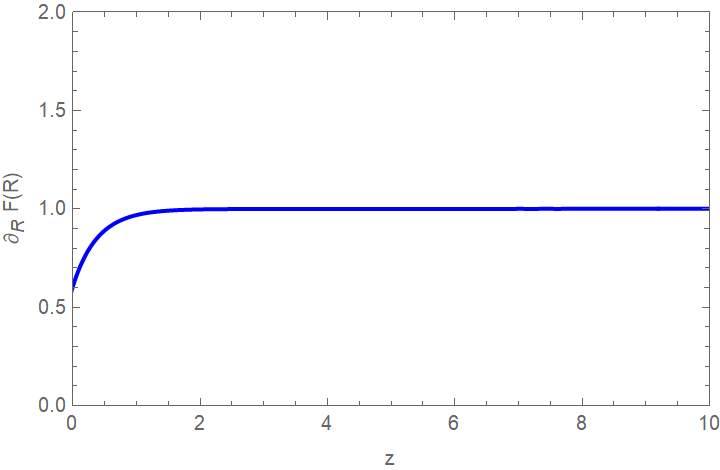}
\includegraphics[width=18pc]{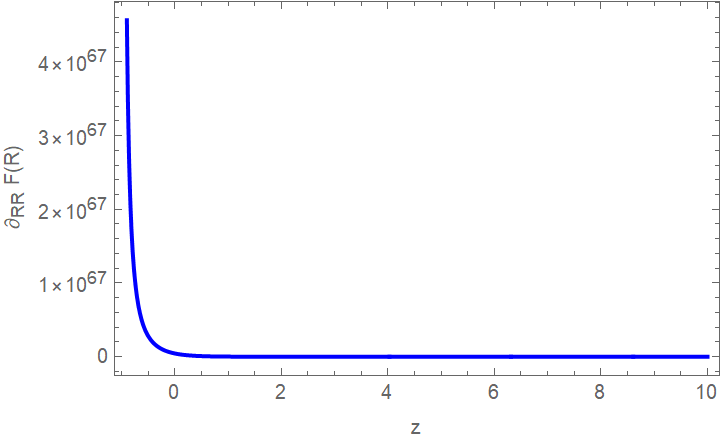}
\caption{The viability criteria (\ref{viabilitycriteria}) for the
model $F(R)=R+\frac{R^2}{M^2}+\lambda  R \exp \left(\frac{\Lambda
\epsilon }{R}\right)-\frac{\Lambda
\left(\frac{R}{m_s^2}\right)^{\delta }}{\gamma }+\lambda \Lambda
\epsilon$ with $\lambda=0.7\times 10^{-3}$, $\gamma=5000$,
$\epsilon=50$ and with $\delta=0.9$.}\label{plotconstrfr2}
\end{figure}

The approach adopted in this paper was finding viable inflationary
$F(R)$ theories on general grounds, starting from the behavior of
the parameter $x= 4\frac{R F_{RRR}}{F_{RR}}$ and focusing on the
inflationary era assuming that this is a slow-roll era. The $F(R)$
gravity function must be a function of the most fundamental mass
scales in cosmology, namely the cosmological constant or the scale
$m_s^2=\frac{\kappa^2 \rho_m^{(0)}}{3}=H_0^2 \Omega_m=1.37 \times
10^{-67} eV^2$, with $\rho_m^{(0)}$ denoting the energy density of
the cold dark matter at the present epoch, or even a scale $M$
which corresponds to the inflationary era and is constrained by
the amplitude of the scalar perturbations. Using this line of
thinking, we developed a framework that enabled us to construct
some general forms of viable $F(R)$ gravity inflationary theories.
Three of the modes we found to be viable were the models of Eq.
(\ref{sol1}), (\ref{modelforde}) and (\ref{simlifiedexponential}),
which we quote here for convenience, so model I is the following,
\begin{equation}\label{sol1exposhow}
F(R)=R+n\Lambda+\frac{1}{32} M^{-2} \left(\Lambda  n (\Lambda  n-8
R) \, \text{Ei}\left(\frac{n \Lambda }{4 R}\right)+4 R
e^{\frac{\Lambda n}{4 R}} (4 R-\Lambda
n)\right)-\frac{\Lambda}{\gamma
}\left(\frac{R}{m_s^2}\right)^{\delta }\, ,
\end{equation}
also,
\begin{equation}\label{modelfordeexposho}
F(R)= R+\frac{R^2}{M^2}+\lambda  R \exp \left(\frac{\Lambda
\epsilon }{R}\right)-\frac{\Lambda
\left(\frac{R}{m_s^2}\right)^{\delta }}{\gamma }+\lambda \Lambda
\epsilon\, ,
\end{equation}
and
\begin{equation}\label{simlifiedexponentialnew}
F(R)=R+\frac{R^2}{M^2}+\lambda
R\,e^{\epsilon\left(\frac{\Lambda}{R}\right)^{\beta}}+\lambda
\Lambda n \epsilon\, ,
\end{equation}
with $\epsilon$, $\lambda$, $\beta$, $n$, $\gamma$ and $\delta$
being dimensionless parameters, and $0<\delta<1$. Our framework
has the remarkable feature of not only providing a framework for
viable inflation, but also provides us with $F(R)$ gravities which
can describe the dark energy era. Note that this is the first time
in the literature that one is able to find a viable $F(R)$ dark
energy era by starting from the requirement of a viable
inflationary era. Of course viable dark energy models which can
also describe inflation in a unified manner also appear in the
literature \cite{Oikonomou:2020qah}, but these models were
constructed by hand on phenomenological basis. In the approach
adopted in this paper, the models emerged by requiring a viable
inflationary era, and as we now show, these models can produce a
viable dark energy era too. Thus our approach provides us with a
framework in the context of which if someone finds a viable
inflationary theory, a simultaneous description of the dark energy
era is achieved with the same model. This section is devoted on
the dark energy era produced by the models (\ref{sol1exposhow})
and (\ref{modelfordeexposho}). Let us review the relevant
formalism for studying the dark energy era for $F(R)$ gravities.
Let us consider $F(R)$ gravity in the presence of perfect fluids,
\begin{equation}
\label{actionde} \centering
\mathcal{S}=\int{d^4x\sqrt{-g}\left(\frac{F(R)}{2\kappa^2}+\mathcal{L}_m\right)}\,
,
\end{equation}
with $\mathcal{L}_m$ standing for the Lagrangian density of the
perfect matter fluids. Let $F(R)$ be in the form of,
\begin{equation}\label{FR}
    F(R)=R+f(R) .
\end{equation}
so upon varying the gravitational action (\ref{actionde}) with
respect to the metric tensor, we get,
\begin{equation} \label{Friedman}
    3 F_R H^2=\kappa^2 \rho_m + \frac{F_R R - F}{2} -3H \dot F_R \, ,
\end{equation}
\begin{equation} \label{Raycha}
    -2 F_R \dot H = \kappa^2 (\rho_m + R_m) + \ddot F -H \dot F \, ,
\end{equation}
with $F_R = \frac{\partial F}{\partial R}$ and the ``dot'' denotes
as usual the derivative with respect to cosmic time. Also $\rho_m$
and $P_m$ denote the matter fluids energy density and  also the
corresponding pressure respectively. The field equations
(\ref{Friedman}),(\ref{Raycha}) can be cast in the form of
Einstein-Hilbert gravity for flat FRW metric as follows,
\begin{equation} \label{Friedtot}
    3 H^2= \kappa^2 \rho_{tot}  \ ,
\end{equation}
\begin{equation} \label{Raychtot}
   -2 \dot H =\kappa^2 (\rho_{tot} + P_{tot}) \ ,
\end{equation}
with $\rho_{tot}$ denoting the total energy density of the total
effective cosmological fluid and $P_{tot}$ denotes the
corresponding total pressure. The cosmological fluid consists of
three parts, the cold dark matter one ($\rho_m$), the radiation
part ($\rho_r$) and the geometric part ($\rho_{DE}$). Hence we
have, $\rho_{tot}=\rho_m + \rho_r + \rho_{DE}$ and also
$P_{tot}=P_m + P_r + P_{DE}$. The geometric fluid drives the
late-time era, and its energy density and effective pressure are,
\begin{equation}\label{rDE}
    \rho_{DE}=\frac{F_R R - F}{2} + 3 H^2 (1-F_R)-3H \dot F_R \, ,
\end{equation}
\begin{equation}\label{PDE}
    P_{DE}=\ddot F -H \dot F +2 \dot H (F_R -1) - \rho_{DE} \, .
\end{equation}
We shall use the redshift
\begin{equation}
    1+z=\frac{1}{a}\, ,
\end{equation}
as a dynamical variable, and also we introduce the statefinder
parameter $y_H (z)$ \cite{Hu:2007nk,Bamba:2012qi,reviews1},
\begin{equation} \label{yhdef}
    y_H(z)=\frac{\rho_{DE}}{\rho_m^{(0)}}=\frac{H^2}{m_s^2}-(1+z)^3-\chi (1+z)^4 ,
\end{equation}
where recall that $\rho_m^{(0)}$ denotes the energy density of the
cold dark matter at the present epoch, and also
$m_s^2=\frac{\kappa^2 \rho_m^{(0)}}{3}=H_0^2 \Omega_m=1.37 \times
10^{-67} eV^2$ and in addition $\chi$ is defined as
$\chi=\frac{\rho_r^{(0)}}{\rho_m^{(0)}} \simeq 3.1 \times
10^{-4}$, with $\rho_r^{(0)}$ being the radiation energy density
at the present epoch.
\begin{figure}
\centering
\includegraphics[width=18pc]{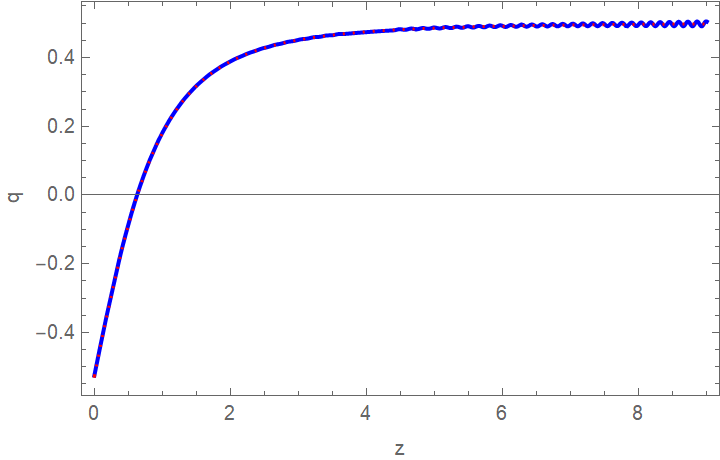}
\includegraphics[width=18pc]{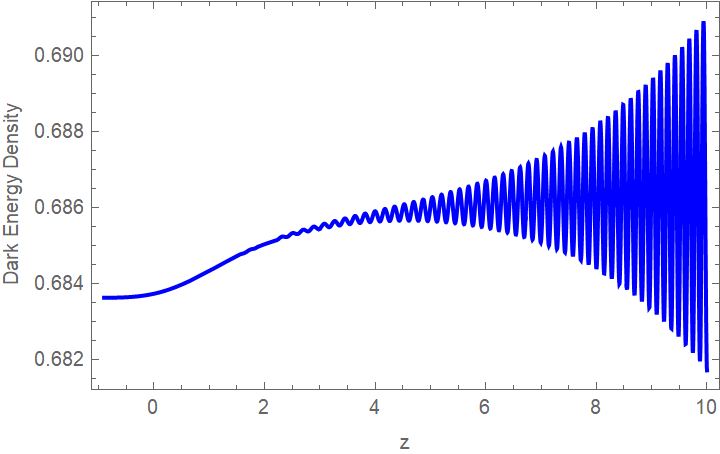}
\includegraphics[width=18pc]{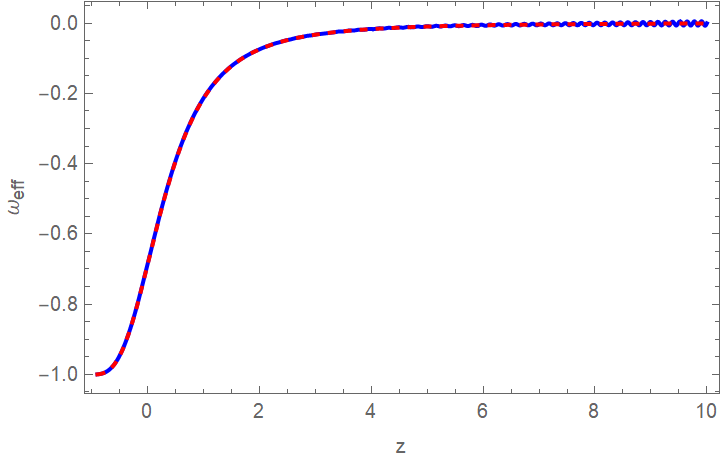}
\caption{Plots of the deceleration parameter $q(z)$ (upper left
plot) the dark energy density parameter $\Omega_{DE}(z)$ (upper
right) and of the total (effective) EoS parameter (bottom plot) as
functions of the redshift for the model
$F(R)=R+n\Lambda+\frac{1}{32} M^{-2} \left(\Lambda  n (\Lambda n-8
R) \, \text{Ei}\left(\frac{n \Lambda }{4 R}\right)+4 R
e^{\frac{\Lambda n}{4 R}} (4 R-\Lambda
n)\right)-\frac{\Lambda}{\gamma
}\left(\frac{R}{m_s^2}\right)^{\delta }$ with $n=0.1$,
$\gamma=0.47\times (0.05)^{\delta}$, and with $\delta=0.001$. The
red curves correspond to the $\Lambda$CDM model.}\label{plotde1}
\end{figure}
Upon combination of Eqs. (\ref{Friedtot}) , (\ref{FR}) and
(\ref{yhdef}), the Friedmann equation can be recast in terms of
the statefinder $y_H$ as follows,
\begin{equation} \label{difyh}
    \frac{d^2 y_H}{dz^2} + J_1 \frac{d y_H}{dz} + J_2 y_H + J_3=0 \, ,
\end{equation}
with the dimensionless functions $J_1$ , $J_2$ , $J_3 $ being,
\begin{equation} \label{J1}
    J_1 = \frac{1}{(z+1)} \Big( -3-\frac{1}{y_H + (z+1)^3 + \chi (z+1)^4}  \frac{1-F_R}{6 m_s^2F_{RR}}\Big) \, ,
\end{equation}

\begin{equation} \label{J2}
    J_2 = \frac{1}{(z+1)^2} \Big( \frac{1}{y_H + (z+1)^3 +\chi (z+1)^4} \frac{2-F_R}{3 m_s^2 F_{RR}}\Big) \, ,
\end{equation}

\begin{equation} \label{J3}
    J_3 = -3(z+1) - \frac{(1-F_R)((z+1)^3 + 2\chi (z+1)^4) + (R-F)/(3 m_s^2)}{(z+1)^2 (y_H + (z+1)^3 +\chi (z+1)^4)} \frac{1}{6 m_s^2 F_{RR}}\, ,
\end{equation}
and in addition $F_{RR}=\frac{\partial^2 F}{\partial R^2}$.
Furthermore, the Ricci scalar is,
\begin{equation}\label{ft10newadd}
R=12H^2-6HH_{z}(1+z)\, ,
\end{equation}
or in terms of $y_H$ we have,
\begin{equation}\label{neweqnrefricciyH}
R(z)=3\,m_s^2\left(-(z+1)\,\frac{d y_H(z)}{dz} + 4 y_H(z) +
(1+z)^3\right)\, .
\end{equation}
We aim to solve Eq. (\ref{difyh}) numerically focusing on the
redshift interval $z=[0,10]$, with appropriate initial conditions.
These are the following, at the final redshift $z_f=10$
\cite{Bamba:2012qi},
\begin{equation}\label{initialcond}
    y_H (z_f) = \frac{ \Lambda}{3 m_s^2} \Big( 1 + \frac{1+z_f}{1000} \Big) \ , \ \frac{d y_H(z)}{dz} \Big |_{z=z_f} = \frac{1}{1000} \frac{ \Lambda}{3 m_s^2},
\end{equation}
with $\Lambda \simeq 11.895 \times 10^{-67} eV^2$. The physical
cosmological quantities in terms of the statefinder $y_H(z)$ are,
\begin{equation}\label{hubblefr}
H(z)=m_s\sqrt{y_H(z)+(1+z)^{3}+\chi (1+z)^4}\, .
\end{equation}
while the Ricci scalar is,
\begin{equation}\label{curvature}
    R(z)=3 m_s^2 \Big( 4 y_H(z) -(z+1) \frac{d y_H (z)}{dz} + (z+1)^3 \Big),
\end{equation}
and in addition, the dark energy density parameter
$\Omega_{DE}(z)$ is,
\begin{equation}\label{OmegaDE}
    \Omega_{DE}(z)=\frac{y_H(z)}{y_H(z)+(z+1)^3 + \chi (z+1)^4}\, ,
\end{equation}
while the dark energy EoS parameter is given by,
\begin{equation}\label{EoSDE}
    \omega_{DE}(z)=-1+\frac{1}{3}(z+1)\frac{1}{y_H(z)}\frac{d y_H(z)}{dz},
\end{equation}
and the total EoS parameter is equal to,
\begin{equation}\label{EoStot}
    \omega_{tot}(z)=\frac{2(z+1)H'(z)}{3H(z)}-1 \, .
\end{equation}
Also the deceleration parameter is defined as,
\begin{equation}\label{declpar}
    q(z)=-1-\frac{\dot H}{H^2}=-1-(z+1)\frac{H'(z)}{H(z)},
\end{equation}
with the ``prime'' denoting differentiation with respect to the
redshift. Finally, the Hubble rate for the $\Lambda CDM$ model is
equal to,
\begin{equation}\label{hubblelcdm}
    H_{\Lambda}(z)=H_0\sqrt{\Omega_{\Lambda} +\Omega_M(z+1)^3 +\Omega_r(z+1)^4 } ,
\end{equation}
with $\Omega_{\Lambda} \simeq 0.68136$ and $\Omega_M \simeq
0.3153$. In addition $ H_0 \simeq 1.37187 \times 10^{-33}$eV is
the Hubble rate at the present epoch according to the latest 2018
Planck data \cite{Planck:2018vyg}. The models we shall use, must
be checked for the redshift interval $z=[0,z_f]$ to see explicitly
whether the constraints \cite{reviews1,Zhao:2008bn},
\begin{equation}\label{viabilitycriteria}
F'(R)>0\,, \,\,\,F''(R)>0\, ,
\end{equation}
holds true for any curvature satisfying $R>R_0$, where $R_0$ is
the present day curvature.
\begin{figure}
\centering
\includegraphics[width=18pc]{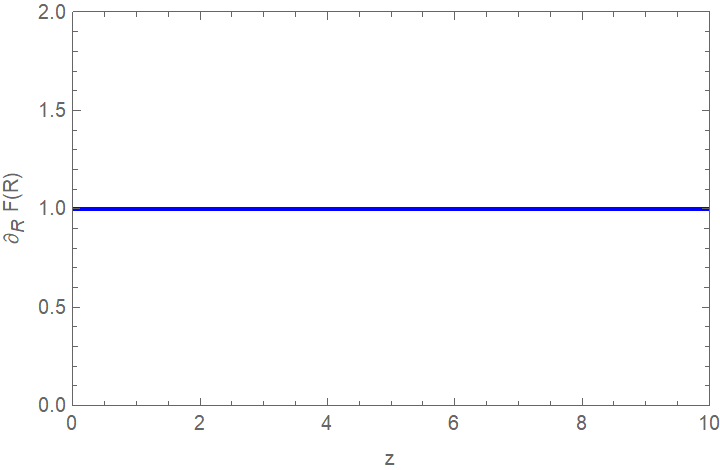}
\includegraphics[width=18pc]{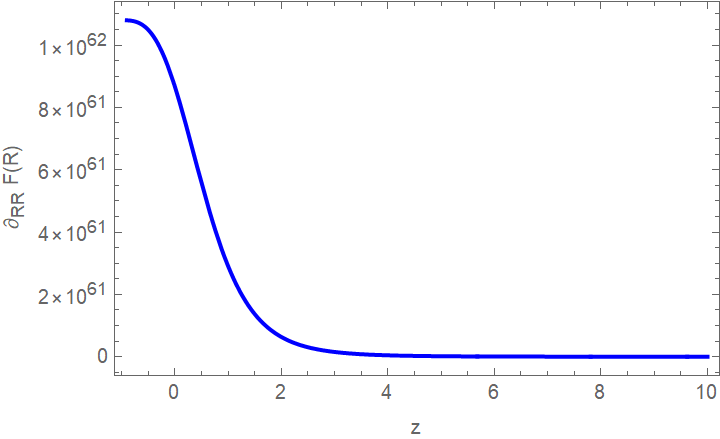}
\caption{The viability criteria (\ref{viabilitycriteria}) for the
model $F(R)=R+n\Lambda+\frac{1}{32} M^{-2} \left(\Lambda n
(\Lambda n-8 R) \, \text{Ei}\left(\frac{n \Lambda }{4 R}\right)+4
R e^{\frac{\Lambda n}{4 R}} (4 R-\Lambda
n)\right)-\frac{\Lambda}{\gamma}\left(\frac{R}{m_s^2}\right)^{\delta}$.}\label{plotconstrfr1}
\end{figure}
Now let us examine the dark energy phenomenology of the models
(\ref{sol1exposhow}), (\ref{simlifiedexponentialnew}) and
(\ref{modelfordeexposho}) in some detail. We shall use the
following numerical values: $M=3.04375 \times 10^{22}$eV which
stems from $M= 1.5\times 10^{-5}\left(\frac{N}{50}\right)^{-1}M_p$
\cite{Appleby:2009uf}, with $N$ being the $e$-foldings number and
for $N\sim \mathcal{O}(50-60)$ we get the value $M=3.04375 \times
10^{22}$eV. Note that this constraint stems from the amplitude of
the scalar perturbations for an $R^2$ inflation theory. Also
$m_s^2=\frac{\kappa^2 \rho_m^{(0)}}{3}=H_0^2 \Omega_m=1.37 \times
10^{-67} eV^2$, with $\rho_m^{(0)}$ denoting the energy density of
the cold dark matter at the present epoch, with
$m_s^2=\frac{\kappa^2 \rho_m^{(0)}}{3}=H_0^2 \Omega_m=1.37 \times
10^{-67} eV^2$. Let us start with the model (\ref{sol1exposhow})
and one example of a viable evolution is obtained by taking
$n=0.1$, $\gamma=0.47\times (0.05)^{\delta}$, with $\delta=0.001$
we obtain $\Omega_{DE}(0)=0.683732$ and $\omega_{DE}(0)=-0.99956$
which are well fitted in the 2018 Planck constraints
$\Omega_{DE}=0.6847 \pm 0.0073$ and $\omega_{DE}=-1.018 \pm
0.031$. In addition we find that $q(0)=-0.525098$ and the total
EoS parameter is $\omega_{tot}(0)=-0.7088$. Also in
Fig.\ref{plotde1} we plot the of the deceleration parameter $q(z)$
(upper left plot) the dark energy density parameter
$\Omega_{DE}(z)$ (upper right) and of the total (effective) EoS
parameter (bottom plot) as functions of the redshift for the model
(\ref{sol1exposhow}) with $n=0.1$, $\gamma=0.47\times
(0.05)^{\delta}$, with $\delta=0.001$, and the red curves
correspond to the $\Lambda$CDM model. We also gathered our results
in Table \ref{table1}. Furthermore, the behavior of $F'(R)$ and
$F''(R)$ for $z=[0,10]$ can be found in Fig. \ref{plotconstrfr1}.
We focused on checking the conditions (\ref{viabilitycriteria}),
namely $F_R > 0$ and $F_{RR} >0$ only in the late-time regime
since it is questionable whether these two hold true at late
times. But during the inflationary regime, the $R^2$ term
dominates and these two conditions automatically hold true. Note
that in order to check this explicitly one must solve the
differential equation (\ref{difyh}) up to redshifts $z\sim
10^{24}$, which correspond to the inflationary regime, but this is
not necessary because at early times the conditions $F_R > 0$ and
$F_{RR} >0$ are automatically satisfied due to the $R^2$ term
leading order domination. As it can be seen in Fig.
\ref{plotconstrfr1} the viability criteria
(\ref{viabilitycriteria}) are satisfied. Thus the model is deemed
as a viable dark energy model, quite similar with the $\Lambda$CDM
model, with the difference that it is describes a dynamical dark
energy era.
\begin{table}[h!]
  \begin{center}
    \caption{\emph{\textbf{Cosmological Parameters Values at present day for the models (\ref{sol1exposhow}) and (\ref{modelfordeexposho}).}}}
    \label{table1}
    \begin{tabular}{|c|c|c|c|c|}
    \hline
      \textbf{Parameter} & (\ref{sol1exposhow})&  (\ref{modelfordeexposho}) & (\ref{simlifiedexponentialnew}) & \textbf{Planck 2018}
      \\  \hline
      $\Omega_{DE}(0)$ & $0.683732$ & $0.685071$ & 0.69019 & $0.6847\pm 0.0073$

 \\  \hline
      $\omega_{DE}(0)$ & $-0.99956$ & $-1.01901$ & -1.036 & $-1.018\pm0.031$

      \\  \hline
      $q(0)$ & $-0.525098$ & $-0.547089$ & -0.57253 & -

      \\  \hline
      $\omega_{tot}(0)$ & $-0.7088$ & $-0.698059$ & -0.68347 & -
      \\  \hline
    \end{tabular}
  \end{center}
\end{table}
Let us now continue with the model (\ref{modelfordeexposho}) and
one example of a viable evolution is obtained by taking
$\lambda=0.7\times 10^{-3}$, $\gamma=5000$, $\epsilon=50$ and with
$\delta=0.9$ we obtain $\Omega_{DE}(0)=0.685071$ and
$\omega_{DE}(0)=-1.01901$ which again are well fitted in the 2018
Planck constraints $\Omega_{DE}=0.6847 \pm 0.0073$ and
$\omega_{DE}=-1.018 \pm 0.031$. Furthermore, we find that
$q(0)=-0.547089$ and the total EoS parameter is
$\omega_{tot}(0)=-0.698059$.
\begin{figure}
\centering
\includegraphics[width=18pc]{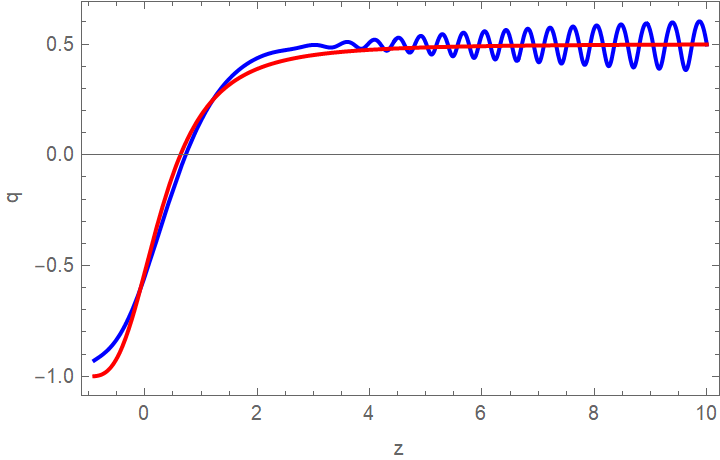}
\includegraphics[width=18pc]{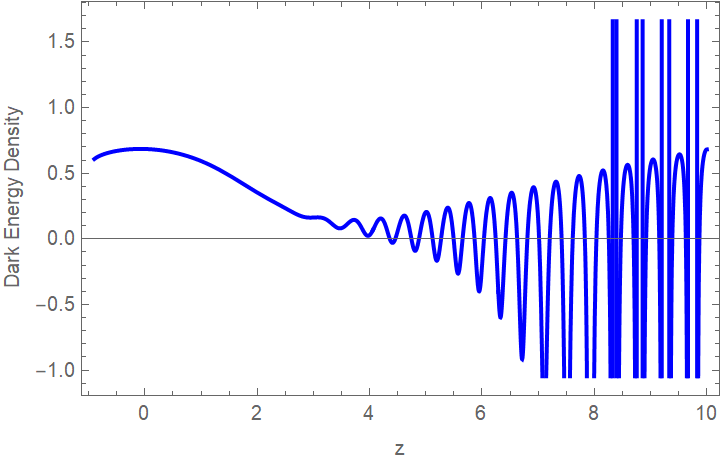}
\includegraphics[width=18pc]{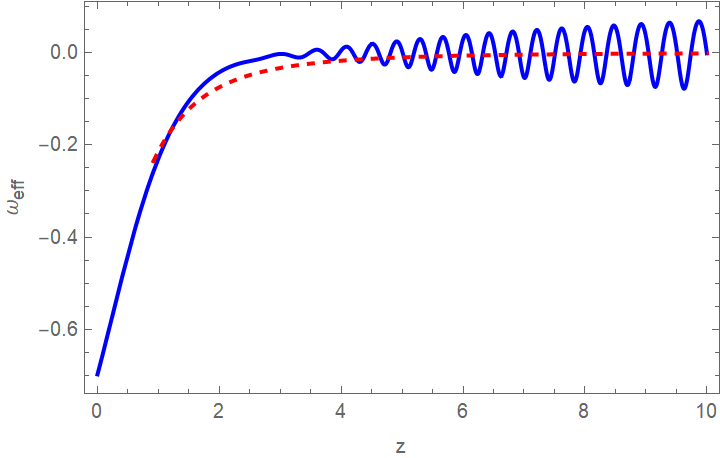}
\caption{Plots of the deceleration parameter $q(z)$ (upper left
plot) the dark energy density parameter $\Omega_{DE}(z)$ (upper
right) and of the total (effective) EoS parameter (bottom plot) as
functions of the redshift for the model
$F(R)=R+\frac{R^2}{M^2}+\lambda  R \exp \left(\frac{\Lambda
\epsilon }{R}\right)-\frac{\Lambda
\left(\frac{R}{m_s^2}\right)^{\delta }}{\gamma }+\lambda \Lambda
\epsilon$ with $\lambda=0.7\times 10^{-3}$, $\gamma=5000$,
$\epsilon=50$ and with $\delta=0.9$. The red curves correspond to
the $\Lambda$CDM model.}\label{plotde2}
\end{figure}
Also in Fig. \ref{plotde2} we plot the of the deceleration
parameter $q(z)$ (upper left plot) the dark energy density
parameter $\Omega_{DE}(z)$ (upper right) and of the total
(effective) EoS parameter (bottom plot) as functions of the
redshift for the model (\ref{modelfordeexposho}) with
$\lambda=0.7\times 10^{-3}$, $\gamma=5000$, $\epsilon=50$ and with
$\delta=0.9$, and the red curves correspond to the $\Lambda$CDM
model. We also gathered our results in Table \ref{table1}.
Furthermore, the behavior of $F'(R)$ and $F''(R)$ for $z=[0,10]$
can be found in Fig. \ref{plotconstrfr2}. As it can be seen in
Fig. \ref{plotconstrfr2} the viability criteria
(\ref{viabilitycriteria}) are satisfied. Thus the model
(\ref{modelfordeexposho}) is deemed as a viable dark energy model,
quite similar with the $\Lambda$CDM model, however, it is notable
that this model exhibits strong dark energy oscillations from
$z\sim 2$ and beyond to higher redshifts.
Let us now study in brief the model
(\ref{simlifiedexponentialnew}) and one example of a viable
evolution is obtained by taking $\lambda=0.8$, $\epsilon=9.1$, and
$n=0.099$ and we obtain $\Omega_{DE}(0)=0.6901$ and
$\omega_{DE}(0)=-1.036$ which again are well fitted in the 2018
Planck constraints $\Omega_{DE}=0.6847 \pm 0.0073$ and
$\omega_{DE}=-1.018 \pm 0.031$. Moreover, we find that
$q(0)=-0.572536$ and the total EoS parameter is
$\omega_{tot}(0)=-0.684673$.
\begin{figure}
\centering
\includegraphics[width=18pc]{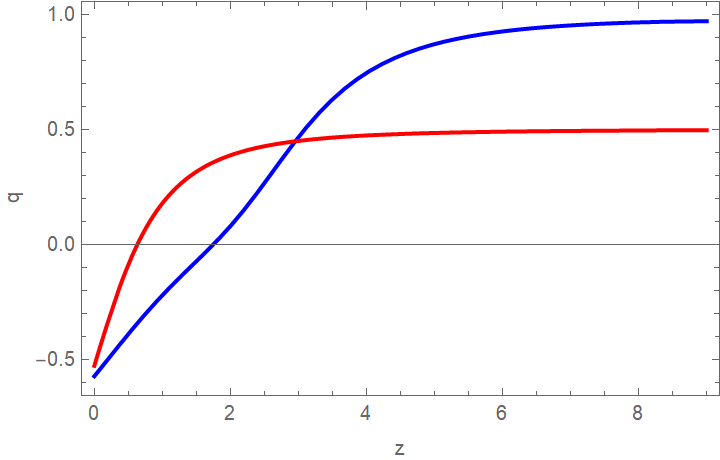}
\includegraphics[width=18pc]{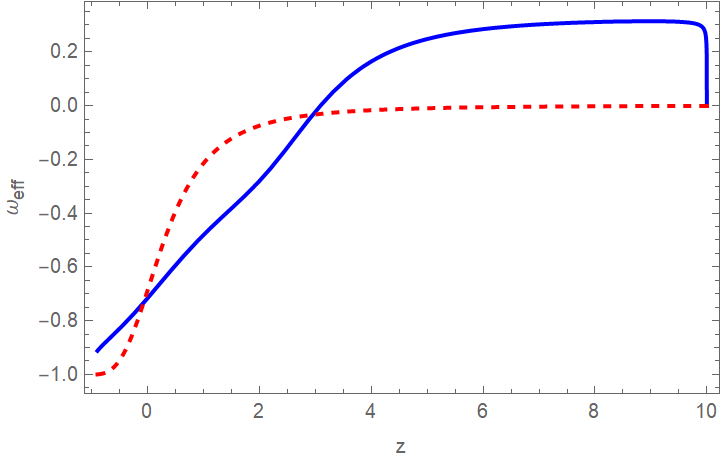}
\caption{Plots of the deceleration parameter $q(z)$ (left plot)
and of the total (effective) EoS parameter (right plot) as
functions of the redshift for the model
$F(R)=R+\frac{R^2}{M^2}+\lambda
R\,e^{\epsilon\left(\frac{\Lambda}{R}\right)^{\beta}}+\lambda
\Lambda n \epsilon$ with $\lambda=0.8$, $\epsilon=9.1$,
$\beta=0.99$ and $n=0.099$. The red curves correspond to the
$\Lambda$CDM model.}\label{plotde2}
\end{figure}
Also in Fig. \ref{plotde2} we plot the of the deceleration
parameter $q(z)$ (upper left plot) the dark energy density
parameter $\Omega_{DE}(z)$ (upper right) and of the total
(effective) EoS parameter (bottom plot) as functions of the
redshift for the model (\ref{simlifiedexponentialnew}) and we also
quote our results in Table \ref{table1}. As it can be seen, there
are significant differences between the model
(\ref{simlifiedexponentialnew}) and the $\Lambda$CDM model,
however the model is a viable model regarding its dark energy
phenomenology.

Thus we demonstrated that our framework provides us with viable
inflationary $F(R)$ gravity models, which simultaneously generate
a viable dark energy era. We need to note that models of this
sort, that contain exponentials of the form $e^{\alpha
\Lambda/R}$, $\alpha>0$ were used in the previous literature based
on a phenomenological approach, but now we demonstrated that these
models originate naturally in viable $F(R)$ gravity inflation
framework. Thus our framework provides a formal and mathematically
rigid unified description of inflation and the dark energy era, by
simply requiring a viable inflationary era for the models, using
the constraints and functional form of the parameter
$x=4\frac{F_{RRR}R}{F_{RR}}$. This is the first time that such
unified framework is formally obtained. We need to note that the
$F(R)$ gravity models studied and presented in this work were
thoroughly examined in \cite{Odintsov:2025jfq} and we confronted
the models with the DESI and other data.

\section{Method for Obtaining $F(R)$ Gravity Inflationary Phenomenology in a Model Agnostic Way}

In this work we aimed to provide some rigid and formal steps
toward finding viable inflationary $F(R)$ gravity models, in a
model agnostic way. To this end, using only the slow-roll
assumption $\dot{H}\ll H^2$ and also the requirement that the
scalaron mass is positive or zero, but also that the scalaron mass
is monotonically increasing function of the Ricci scalar in the
large curvature limit, or even zero to capture the extremum case,
and we ended up in the following results:
\begin{itemize}
    \item The inflationary phenomenology of any $F(R)$ gravity
    theory is determined by two parameters, the parameter $x=\frac{R
    F_{RRR}}{F_{RR}}$and the first slow-roll index
    $\epsilon_1=-\frac{\dot{H}}{H^2}$. If these are calculated at
    the first horizon crossing, the phenomenology of an arbitrary
    $F(R)$ gravity model can be determined, with the relations $n_s-1=-4\epsilon_1+x\epsilon_1$ and also $r\simeq \frac{48
    (1-n_s)^2}{(4-x)^2}$.
    \item The general $F(R)$ gravity must be a function of the
    fundamental scales of cosmology, the cosmological constant
    $\Lambda$ and of $m_s^2$ related to the energy density of cold
    dark matter at present day. So the $F(R)$ gravity must be a
    function of $F(R,\Lambda,m_s^2)$.
    \item Remarkably, the viable inflationary $F(R)$ gravity models
    also generate a viable dark energy era, and interestingly
    enough these contain exponentials of the form $e^{\Lambda/R}$
    and are deformations of $R^2$ gravity.
     \item Viable inflationary $F(R)$ gravity models yield $-1\leq x\leq 0$ and
also the parameter $y=\frac{RF_{RR}}{F_R}$ for these models is in
the range $0<y\leq 1$, during the slow-roll inflationary era.

     \item Any model with $x\gg 1$ during the slow-roll era is non-viable regarding
     inflation.

      \item Any model with $x<-1$ or $x>1$ violates the de Sitter criterion and also yields a large spectral index, if the first slow-roll
      index is of the order $\epsilon_1\sim \mathcal{O}(10^{-3})$ as dictated by the Planck data \cite{Planck:2018jri}, and thus is non viable.

\end{itemize}
Thus, in order to perform and study inflationary dynamics in our
framework, one needs the values of $x$ and of the first slow-roll
index at first horizon crossing. In some cases for which $x$ can
be evaluated, for example when $x\sim 0$ or even when $x=const$,
one needs only the value of the first slow-roll index at first
horizon crossing. At this point, let us provide a very simple
approximate technique in order to have an approximate value for
the first slow-roll index, without the details of the Hubble rate
needed. All that is needed is the slow-roll approximation, so
$\dot{H}\ll H^2$ and also the functional form of the $F(R)$
gravity. Let us start with the Friedmann equation for $F(R)$
gravity which is,
\begin{align}
\label{JGRG15gfhgfyg} 0 =& -\frac{F(R)}{2} + 3\left(H^2 + \dot
H\right) F_R(R) - 18 \left( 4H^2 \dot H + H \ddot H\right)
F_{RR}(R)\, ,
\end{align}
with $F_{RR}=\frac{\mathrm{d}^2F}{\mathrm{d}R^2}$, and in addition
the Ricci scalar for the FRW metric is $R=12H^2 + 6\dot H$. During
the slow-roll era, when $\dot{H}\ll H^2$, the Friedmann equation
is written,
\begin{equation}\label{approximate}
3 H^2F_R-\frac{F(R)}{2}-72H^2\dot{H}F_{RR}\sim 0\, ,
\end{equation}
so for $R\sim 12 H^2$, we get,
\begin{equation}\label{approximateepsilon1}
\epsilon_1\sim \frac{2 F-F_R R}{2F_{RR}R^2}\, .
\end{equation}
We used this formula to obtain an approximate value for the first
slow-roll index for the case of a pure power-law $F(R)$ gravity
which leads to a constant $x$ parameter, see for example Eq.
(\ref{epsilonleading}). So finding the first slow-roll index
during inflation using formula (\ref{approximateepsilon1}) and
also knowing $x$, one may have a concrete idea about the viability
of a given $F(R)$ gravity model. Let us give an example here to
validate our findings, using the well-known $R^2$ model. For
$F(R)=R+\frac{R^2}{M^2}$, where $M= 1.5\times
10^{-5}\left(\frac{N}{50}\right)^{-1}M_p$ \cite{Appleby:2009uf},
with $N$ being the $e$-foldings number during inflation. This
value of $M$ is obtained by using the constraint on the amplitude
of the scalar perturbations for the $F(R)$ gravity model. So for
$N\sim \mathcal{O}(50-60)$ we have $M=3.04375 \times 10^{22}$eV.
Using the approximation (\ref{approximateepsilon1}), the first
slow-roll index reads,
\begin{equation}\label{firstr2model}
\epsilon_1=\frac{M^2}{4 R}\, ,
\end{equation}
so for $M=3.04375 \times 10^{22}$eV and for $H_I\sim
10^{14}\,$GeV, we approximately have $\epsilon_1\sim 0.0082$ which
is quite close to the values $\epsilon_1\sim 1/(2N)$ obtained
analytically for the $R^2$ model. Thus this method enables us to
obtain at least the order of magnitude of the first slow-roll
index and decide whether a given arbitrary model of $F(R)$ gravity
can be viable. However, for more concrete results, one needs to
evaluate the first slow-roll index numerically, which can be
demanding. The optimal feature of our method is that only the
first slow-roll index must be evaluated. Thus one needs to
numerically solve the Friedmann and Raychaudhuri equations for
appropriate initial conditions and determine the first slow-roll
index. An estimate of the order of the first slow-roll index may
provide useful feedback for potentially viable inflationary $F(R)$
gravity models.

The method can be summarized in the following steps:
\begin{enumerate}
    \item Select an $F(R)$ gravity model and evaluate the
    parameter $x$ during the inflationary era using the slow-roll assumption. Then evaluate
    approximately the first slow-roll index $\epsilon_1$ using the
    approximation (\ref{approximateepsilon1}). If $-1\leq x\leq 0$ and the
    de Sitter criterion applies, and in addition if $\epsilon_1\ll
    1$, then the model is possibly viable. If $x\sim 0$, and the dominant terms during the slow-roll era is an $R^2$ term, then the
    model is certainly viable and it is an $R^2$ deformation. In
    both cases, the inflationary observational indices are given
    by $n_s-1=-4\epsilon_1+x\epsilon_1$ and also $r\simeq \frac{48
    (1-n_s)^2}{(4-x)^2}$.
    \item If $\epsilon_1\gg 1$ or $\epsilon_1\sim \mathcal{O}(1)$
    during the slow-roll era, then the model is not viable. This is because if $x\gg 1$ (in fact if $x\sim \mathcal{O}(10)$), the slow-roll parameter $\epsilon_4$ in
    Eq. (\ref{epsilon4finalnew}) would be larger than unity and thus the slow-roll approximation would be violated. Thus these models with large $x$, that is,
    large enough to break the perturbative expansion in terms of the slow-roll parameters, would violate the slow-roll conditions and thus would be rendered
    non-viable, regarding their inflationary dynamics.
    \item Suppose that a viable model is found with $ x\leq 0$, and also $\epsilon_1\ll 1$ (much more small than $\mathcal{O}(10^{-3})$),
    if one needs more precision then one must evaluate numerically
    the first slow-roll index solely.
    \item It is not certain that if a model yields $-1\leq x\leq
    0$ will provide a viable inflationary phenomenology, but all
    the models which provide a viable phenomenology yield $-1\leq x\leq
    0$. This is clearly an attractor behavior among the $F(R)$
    gravity models which unify inflation and the dark energy era.
\end{enumerate}
Thus our method provides certain results if a model is non-viable,
and also can yield substantial evidence on whether a model is
viable or not. Also in the cases of simple $R^2$ deformations, the
model is certainly viable regarding its inflationary
phenomenology. This is a solid step toward viable $F(R)$ gravity
inflationary phenomenology modelling. However, special caution is
needed for $F(R)$ gravity models which lead to power-law
evolutions which yield $\dot{\epsilon}_1=0$. These models will be
dealt in a separate article.

In our approach to the problem we aimed to present some attractor
properties in $F(R)$ gravity which point towards a model agnostic
formulation of $F(R)$ gravity. This is a very difficult task, but
with our analysis based on a simple slow-roll assumption and the
de Sitter scalaron mass criterion, we simplified the formalism of
$F(R)$ gravity inflation to simply finding the value of the first
slow-roll index. This by itself is simplifying significantly the
analysis of general $F(R)$ gravity inflation.

There are cases though that it is hard to simply calculate the
first slow-roll index on simple grounds due to lack of
analyticity, one cannot simply calculate it. In fact the simplest
and the only case known is the $R^2$ inflation. We provided the
formula (\ref{approximateepsilon1}) for an estimation of the first
slow-roll index, which in the case of $R^2$ gravity gives a value
quite close to the analytic evaluation of the index.

For complicated cases a numerical analysis is needed unavoidably.
However this is not a subject that can be developed in a simple
section of an article. It rather deserves an article focusing on
the numerical evaluation itself. Such a task is quite difficult,
and some fist steps in the context of other theories are done in
\cite{Pozdeeva:2025ied,Pozdeeva:2024ihc,Pozdeeva:2021iwc}. But the
numerical analysis aims in solving the field equations using
appropriate initial conditions. Thus the big question for
inflation is the choice of initial conditions. This is not a
simple task because one is not certain whether all the initial
conditions lead to de Sitter solutions, for arbitrary $F(R)$
gravity. In fact it was shown in \cite{Odintsov:2017tbc} and in
section III in this paper, that the de Sitter subspace of $F(R)$
gravity contains two kinds of fixed points, a stable de Sitter and
an unstable de Sitter. Thus, depending on the initial conditions
one may end up to the stable or unstable de Sitter, if and only if
one achieves to reach the de Sitter subspace of the total $F(R)$
gravity by using the appropriate $F(R)$ gravity function and the
appropriate initial conditions.

Hence, the numerical analysis requires a focused research article
devoted to this non-trivial problem. With the numerical analysis
one must ensure the following, given a specific form of $F(R)$
gravity: firstly, that a de Sitter or quasi-de Sitter evolution is
achieved, or an inflationary power-law evolution, or some
inflationary evolution in general, and secondly that this era is a
slow-roll era that guarantees at least 60 $e$-foldings. In order
to achieve these tasks, one must find the appropriate initial
conditions, and this is one of the most difficult tasks in
inflationary cosmology. The assumption of an initial Bunch-Davies
vacuum at first horizon crossing, is of course an assumption, a
working example. After these tasks are appropriately amended ,
then one may proceed to the evaluation of the first slow-roll
index for a given $F(R)$ gravity function.

Thus with our article, we paved the way towards a model agnostic
formalism in $F(R)$ gravity inflation. The next step is performing
a numerical analysis, which requires an entire new article and
thorough study, which we defer for the future.

\section{Conclusions}

In this work we aimed to provide a theoretical framework that will
enable the study of $F(R)$ gravity in a model-independent way. The
focus was to provide formulas that will determine in a formal way
whether a class of models generates a viable inflationary era or
not. Also we investigated the general form of $F(R)$ gravity that
will be able to describe inflation and dark energy in the same
theoretical framework, from first principles, without adding by
hand terms, or choosing a convenient $F(R)$ gravity from the
beginning. Our findings are successful since we derived several
criteria that a viable $F(R)$ gravity inflationary theory must
satisfy and in addition, the viable models remarkably lead to a
simultaneous successful description of the dark energy era.

Starting from first principles, an $F(R)$ gravity that will be
able to describe both inflation and the dark energy era, must
depend on the cosmological constant $\Lambda$, the mass scale
$m_s^2$ related to the current energy density of cold dark matter
and probably on a mass scale constrained by the amplitude of the
scalar perturbations. However, the fundamental scales are solely
$\Lambda$ and $m_s^2$. After that, assuming that a slow-roll era
is realized, thus $\dot{H}\ll H^2$ and also that the first
slow-roll index satisfies $\dot{\epsilon}_1\neq 0$, we formulated
the slow-roll inflation in the context of a general $F(R)$ gravity
and we derived the observational indices of inflation. Also by
requiring that the scalaron mass is a monotonically increasing
function of $R$, or even that it has an extremum, this proved to
have dramatic consequences for the allowed $F(R)$ gravities, since
if $m^2(R)$ is a monotonically increasing function, this means
that the scalaron mass is small at small curvatures and large at
large curvatures, which is theoretically motivated by the
late-time behavior of the scalaron mass. Indeed if a unified
description of inflation and of the dark energy is needed, then we
need the theory to have a large scalaron mass primordially, while
at late times the scalaron mass must be small. We examined several
models of interest and indicated the features of viable and
non-viable $F(R)$ gravity models. We also highlighted the
importance of exponential deformations of the $R^2$ model of the
form,
$$
F(R)=R+\frac{R^2}{M^2}+\lambda
R\,e^{\epsilon\left(\frac{\Lambda}{R}\right)^{\beta}}+\lambda
\Lambda n \epsilon
$$
which stem naturally by the formalism developed in this paper and
these models provide a unified description of inflation and dark
energy.

We also examined the constant $x$ case and we demonstrated that
the standard literature approach for power-law $F(R)$ gravities,
which relate the models to power-law evolution, is wrong. As we
demonstrated the power-law $F(R)$ gravity models are capable of
providing viable inflation. Our formalism can also give a hint on
the values of the first slow-roll index during the inflationary
era, using the approximate formula,
$$
\epsilon_1\sim \frac{2 F-F_R R}{2F_{RR}R^2}\, ,
$$
which is derived by the Friedmann equation, using only the
slow-roll approximation. Thus our method makes the study of $F(R)$
gravity inflation quite easy since only the parameter $x$ and the
first slow-roll index are needed for the analysis. The analysis
can be strengthen if one evaluates numerically the first slow-roll
index. Notably, the viable $F(R)$ gravity inflationary theories
are either deformations of $R^2$ or $\alpha$-attractor-like
theories. More importantly, the viable $R^2$ deformations provide
simultaneously an also viable dark energy era, compatible with the
latest Planck data and also similar to the $\Lambda$CDM model. In
conclusion, our main results are the following:
\begin{itemize}
    \item All the viable $F(R)$ gravity models which can describe
    simultaneously inflation and the dark energy, in a unified way, yield a parameter $x$
in the range $-1\leq x\leq 0$
    \item For the viable unification models, the de Sitter scalaron mass is small and positive at late times,
and large and positive at early times.
\end{itemize}
One task we did not perform in this work, is the analysis of
$F(R)$ gravity models which lead to a constant first slow-roll
index, namely $\dot{\epsilon}_1=0$. These models cannot be
described by the formalism developed in this work and will be
studied in a future work. In addition, it is tempting to consider
chameleon $F(R)$ gravity effects in the context of our viable
$F(R)$ gravities, since the  scalaron mass is large for large
curvatures. Thus in strong gravity regimes, such as near compact
objects, chameleon effects might be important. Chameleon $F(R)$
gravities frequently appear in the literature
\cite{Brax:2008hh,Numajiri:2023uif,Katsuragawa:2019uto}, thus it
is tempting to revisit the above research line in the context of
our unified $F(R)$ gravity models. Finally, the study of the
reheating era for the models that provide a unified description of
the early and late-time eras is also compelling, since the
reheating era will be affected by the same terms which affect the
dark energy era. We hope to address some of these issues in a
future work.

\section*{Acknowledgments}

This research has been is funded by the Committee of Science of
the Ministry of Education and Science of the Republic of
Kazakhstan (Grant No. AP26194585) (V.K. Oikonomou).

\end{document}